\documentclass{jpp}
\usepackage{graphicx}

\usepackage[T1]{fontenc}
\usepackage[utf8]{inputenc}
\usepackage{amsmath}
\usepackage{color}
\usepackage{mathtools}
\usepackage{amssymb}
\usepackage{hyperref}
\usepackage{natbib}
\usepackage{subcaption}

\newcommand{\dd}[1]{\mathrm{d}#1 \,}

\newcommand{\pdev}[2]{\frac{\partial #1}{\partial #2}}

\newcommand{\dev}[2]{\frac{\mathrm{d}#1}{\mathrm{d}#2}}
\renewcommand{\v}[1]{\boldsymbol{#1}}

\newcommand{\crl}[1]{\langle #1 \rangle}
\newcommand{\delgam}{\Delta\rmGamma}

\newcommand{\addedit}[1]{\textcolor{red}{#1}}

\shorttitle{Universal Lynden-Bell equilibria}
\shortauthor{R. J. Ewart et al.}

\title{Non-thermal particle acceleration and power-law tails via relaxation to universal Lynden-Bell equilibria}

\author{R. J. Ewart\aff{1,2} \corresp{\email{robert.ewart@physics.ox.ac.uk}}, M. L. Nastac\aff{1,3}, \and A. A. Schekochihin\aff{1,4}}

\affiliation{\aff{1} Rudolf Peierls Centre for Theoretical Physics, University of Oxford, Oxford, OX1 3PU, UK
\aff{2}Balliol College, Oxford, OX1 3BJ
\aff{3}St John's College, Oxford, OX1 3JP
\aff{4}Merton College, Oxford, OX1 4JD
}

\begin{document}

\maketitle
\begin{abstract}
Collisionless and weakly collisional plasmas often exhibit non-thermal quasi-equilibria. Among these quasi-equilibria, distributions with power-law tails are ubiquitous. It is shown that the statistical-mechanical approach originally suggested by \cite{LyndenBell67} can easily recover such power-law tails. Moreover, we show that, despite the apparent diversity of Lynden-Bell equilibria, a generic form of the equilibrium distribution at high energies is a `hard' power-law tail $\propto \varepsilon^{-2}$, where $\varepsilon$ is the particle energy. The shape of the `core' of the distribution, located at low energies, retains some dependence on the initial condition but it is the tail (or `halo') that contains most of the energy. Thus, a degree of universality exists in collisionless plasmas.
\end{abstract}
\section{Introduction}
It is well known that the ultimate fate of a homogeneous collisional plasma is to become a Maxwellian. This result was first inferred for neutral particles by \cite{Maxwell1860} on statistical grounds and given solid dynamical foundation by \cite{Boltzmann} with his collision integral. Plasma physics was to wait for \cite{Landau1936} and later \cite{Balescu60} and \cite{Lenard60} to be equipped with its own collision integral, and the resulting universality. Nevertheless, distributions with power-law tails, a far cry from Maxwellian equilibria, are observed in a myriad of plasma systems including cosmic rays \citep{Becker_Tjus_2020,amato_casanova_2021}, the solar corona and solar flares \citep{Dudik_2017, Oka_2018}, the solar wind \citep{Gloeckler2008,Fisk_2014,Livadiotis_2018,Moncuquet_2020}, the Earth's magnetosheath \citep{Birn_2012, Ergun_2020}, and laser plasmas \citep{Rigby_2018,Hartouni_2022}. 

That such non-Maxwellian distributions emerge should perhaps come as no surprise. In a plasma, the timescale of relaxation to Maxwellian equilibrium is associated with two-body Coulomb collisions, but, due to the long-range nature of the forces involved, the plasma may evolve, exchanging energy between fields and particles, on much shorter timescales. Indeed, in the absolute absence of collisions, the Vlasov equation has an infinite set of nonlinearly stable equilibria: all distributions that are monotonically decreasing functions of particle energy are certainly stable \citep{Gardner63}. However, the set of all monotonically decreasing functions of energy is very large, and is certainly not exhausted by the Maxwellian equilibrium, which depends on only two parameters. It is, therefore, an outstanding challenge to determine whether any of these stable collisionless equilibria are naturally favoured by the dynamics of the system in a way that is not sensitive to initial conditions, i.e., whether a degree of universality exists in collisionless plasmas. It is certainly the case that nature appears to prefer distributions with power-law tails, and direct numerical simulations have indicated that power-law tails are the natural result of a number of dynamical processes including relativistic and non-relativistic shocks \citep{Sironi_2010,Caprioli_2014,Crumley_2019}, magnetic reconnection \citep{Sironi_2014,werner_uzdensky_2021, uzdensky_2022}, and various types of plasma turbulence \citep{Kunz_2016, Comisso_2018, Comisso_2022,Zhdankin_2017,Zhdankin_2019,Zhdankin_2021b,Zhdankin_2022}.

In addition to suggesting dynamical paths towards distributions with power-law tails, there have been multiple attempts to justify the ubiquity of such distributions from a thermodynamic point of view. This is, however, entangled with the question of whether the standard Gibbs--Shannon entropy is applicable to systems with long-range interactions and, if not, then what entropy should be used. Naturally, many entropies have emerged to fill this niche. A popular contender is the \cite{Tsallis1988} entropy (or $\alpha$-structural entropy: see \citealt{Havrda1967}). The Tsallis entropy was  designed to be a non-additive version of the Gibbs--Shannon entropy as a way to model systems with correlations that therefore should be non-extensive (see, e.g., \citealt{Livadiotis_2009, Pierrard_2010} and references therein). While this model produces good fits to observed distributions, it has a free parameter that is needed to quantify the degree of the non-extensivity and cannot be determined without fitting data, or additional input of physics currently lacking (note some recent progress suggesting that this additional physics might be deducible from free-energy considerations: \citealt{Zhdankin_2021a,Zhdankin_2022}). 

An early attempt to tackle the question of entropy in collisionless systems was made by \cite{LyndenBell67}. Let us consider a system of $N$ particles with canonical positions~$\v{r}_{i}$ and momenta~$\v{p}_{i}$ that evolve subject to a Hamiltonian~$\mathcal{H}(\v{r}_{i},\v{p}_{i})$. Such a system can be said to be `collisionless' if the evolution equation for the single-particle distribution function~$f(\v{r},\v{p})$ is well approximated by an effective Hamiltonian~$\mathcal{H}^{\mathrm{eff}}(\v{r},\v{p})$ acting on a single particle (i.e., if the mean-field dynamics are a sufficiently good approximation to the true dynamics), viz., 
\begin{equation}
\label{Eqn:S1:E1}
\pdev{f}{t} + \pdev{\mathcal{H}^{\mathrm{eff}}}{\v{p}}\cdot\pdev{f}{\v{r}} - \pdev{\mathcal{H}^{\mathrm{eff}}}{\v{r}}\cdot\pdev{f}{\v{p}} = 0.
\end{equation}
In his original treatment, Lynden-Bell focused on relaxation of stellar systems, but the spirit of his statistical mechanics is the same for all collisionless systems, including plasmas. While keeping the calculations as general as possible, one can think of (\ref{Eqn:S1:E1}) as the collisionless Vlasov equation for a plasma, which could be electrostatic or electromagnetic in a non-relativistic or relativistic regime. The collisionless dynamics described by~(\ref{Eqn:S1:E1}) conserve an infinite number of invariants, equivalent to conserving the volume of level sets of the distribution function $f(\v{r},\v{p})$ in phase space. Thus, the dynamics can be viewed as an extremely complicated rearranging of the elements of phase space, which, however much they are distorted and stirred, will keep the same level sets (often referred to as `waterbags', in analogy with parcels of incompressible fluid). Lynden-Bell posited that, after a short time, the exact phase-space density~$f(\v{r},\v{p})$ would become so chaotic that it could be treated as a random field and that any measurement of it---in practice, of a coarse-grained version of it---was in fact a measurement of the mean phase-space density. This allowed the construction of a statistical mechanics, with an entropy closely related to the Gibbs--Shannon entropy, that encoded an infinite number of invariants and thus predicted the steady states from a given initial condition. These steady states are the Lynden-Bell equilibria.

Since its genesis, Lynden-Bell's theory (often referred to as the theory of `violent relaxation') has received continued attention both thermodynamically \citep{Chavanis_1996,Arad_2005,Chavanis_2006a,Chavanis_2006b,Levin_2008,Levin_2014} and dynamically, viz., effective `collisionless collision integrals' have been proposed that recovered Lynden-Bell equilibria as their fixed points \citep{Kadomtsev_Pogutse70,Severne_1980,Chavanis2004,Chavanis_2021,Ewart_2022}. However, the main strength of the theory is also its weakness. Unlike in the non-extensive entropy formulations, there is no \textit{ad-hoc} parameter in the Lynden-Bell theory: equilibria are uniquely determined by the `waterbag content' of the initial conditions. However, this necessarily means that the equilibria depend (seemingly, in a complicated way) on an infinite family of invariants (sometimes referred to as `Casimirs'). This has limited any actual calculations with Lynden-Bell equilibria to simplified situations with only a small number of level sets (in practice, between one and three, e.g., \citealt{Assllani_2012}). At any rate, the intricate dependence on an infinite family of invariants might not appear to be a step towards general power-law tails or any other meaningful form of universality. 

To see just how non-universal Lynden-Bell equilibria can be, one only needs to consider the relation between the Lynden-Bell equilibria and the aforementioned Gardner distributions. Should the initial distribution be a monotonically decreasing function of particle energy, then it is a Gardner distribution and there are no possible rearrangements of the phase volume that do not increase energy. Hence the only state available via collisionless dynamics is this Gardner distribution, which must therefore be its own Lynden-Bell equilibrium\footnote{This takes a surprising amount of work to show formally: see Appendix \ref{Appendix: Degenerate Lynden-Bell equilibria}.}. But since \textit{any} monotonically decreasing function of energy is a Gardner distribution, these minimum-energy states are clearly highly non-universal. However, this is only a good intuition for systems where the number of level sets is small or where phase-volume conservation conspires with energy conservation to render much of the phase space inaccessible to the system (as is the case for Gardner distributions). In this paper, by solving for the full Lynden-Bell equilibria numerically (as well as analytically, in a tractable limit), we will show that most Lynden-Bell equilibria are much more generic. Namely, we will show that, in the limit of a continuum of level sets, and for energies sufficiently greater than the ground-state energy (the energy of the corresponding Gardner distribution), the Lynden-Bell equilibria exhibit power-law tails at high energies, typically with a scaling of $\varepsilon^{-2}$, where $\varepsilon$ is the particle energy. 

The physical argument for these power-law tails is as follows. Phase-volume conservation effectively makes the particles occupying each waterbag (level set of the distribution) behave as if they were members of a separate species, which can communicate with the other waterbags only via the equilibration of some effective `temperature' subject to competition for the same volumes of phase space. In essence, this turns the system into a ensemble of many different fermionic species, all of which exclude each other. When the system is in its minimum-energy (ground) state, the Gardner distribution, the competition for phase volume is the overpowering factor, giving the distribution a highly non-universal shape. However, when the energy of the system is increased, more of the phase space becomes accessible, so, as in Fermi--Dirac statistics, the competition for the same phase volume becomes weaker. For sufficiently large energies, the competition for any volume of phase space is minuscule. In this limit, each waterbag will form its own Maxwellian distribution, in thermal equilibrium with all other waterbags. However, despite these Maxwellian equilibria having the same effective temperature, they will have different thermal spreads because waterbags of larger phase-space density `cost' more energy to be placed at a given momentum $p$ in phase space. The true distribution function is recovered by summing up (in the limit of many waterbags, integrating) the contributions from each of these Maxwellians to the mean phase-space density. This procedure naturally gives rise to a power-law tail that depends on the relative weighting for each Maxwellian (this is qualitatively similar to the formalism of `superstatistics'; cf. \citealt{Beck_2003,Chavanis_2006b,Davis_2023}). This weighting turns out to depend only weakly on the level sets of the initial condition. For a wide class of initial conditions, the resulting Lynden-Bell equilibria turn out to have the same universal power-law tail, $\varepsilon^{-2}$\footnote{This is similar to how Zipf's law arises in systems where one marginalises over a `hidden variable' (cf. \citealt{Mora_2011,Schwab_2014,Aitchison_2016}).}.

The rest of this paper is organised as follows. In Section~\ref{Section:2}, we will review briefly the Lynden-Bell formalism, to state the problem and establish notation. We will then proceed to perform a systematic exploration of the nature of the Lynden-Bell equilibria. In Section~\ref{subsection:Lynden-Bell equilibria as excited Gardner distribution functions}, we will argue that to each initial condition, one can uniquely assign a Gardner distribution function with the same Casimir invariants (waterbag content). All Lynden-Bell equilibria can thus be viewed as the result of adding some amount of energy to a Gardner distribution with the same waterbag content and letting it reach a maximum-entropy state (one can think of this approach as describing how adding energy to a population of collisionless particles causes them to form a `non-thermal' distribution with a tail). In Section~\ref{subsection:Waterbag content of Gardner distributions}, we will show that the function describing the waterbag content of a large class of Gardner distributions has a relatively generic form, which will contribute to the universality of the resulting equilibria. In Section~\ref{subsection:Non-degenerate Lynden-Bell equilibria}, we will solve for the Lynden-Bell equilibria in the limit where the energy of the system far exceeds the energy of the corresponding Gardner distribution. This will ensure that competition for volumes of phase space can be neglected. This makes the problem analytically tractable, and the resulting analytical solution will exhibit the universal power-law tail $\propto \varepsilon^{-2}$ at high energies. In Section~\ref{Section:Numerical result}, by solving for the Lynden-Bell equilibria numerically, we will show that the qualitative features of this analytical solution are retained even for energies that are of the same order as the energy of the Gardner distribution. Therefore, a large class of Lynden-Bell equilibria display a universal power-law tail. This tail contains much of the distribution's energy, whereas the low-energy `core' retains some dependence on the initial conditions. In Section~\ref{Section:Conclusion}, we summarise our findings and discuss their implications for real (observed) plasmas.

\section{Lynden-Bell's statistical mechanics}
\label{Section:2}
In this section, we present a brief re-derivation of  Lynden-Bell's equilibria as applied to a homogeneous system. We begin with the collisionless Vlasov equation~(\ref{Eqn:S1:E1}) evolving a single species of particles. As well as particle number, momentum, and total energy (i.e., of fields and particles),~(\ref{Eqn:S1:E1}) conserves an infinite number of `Casimir' invariants, e.g., the volume of phase space where the exact phase-space density is greater than a given value $\eta$:
\begin{equation}
\label{Eqn:S2:E1}
\rmGamma(\eta) = \iint\dd{\v{r}}\dd{\v{p}}H(f(\v{r},\v{p}) - \eta) = \mathrm{const},
\end{equation}
where $H(x)$ is the Heaviside function (unity for~$x>0$ and zero otherwise). As discussed above, despite the existence of these invariants, the system's evolution can still be highly chaotic, which prompted Lynden-Bell to consider the exact phase-space density~$f(\v{r},\v{p})$ as a random field. Therefore, one may introduce the probability density~$P(\v{r},\v{p},\eta)$ for the exact phase-space density~$f(\v{r},\v{p})$ to take the value~$\eta$ at position~$(\v{r},\v{p})$ \citep{Robert_Sommeria_1991,Chavanis_1996}. The distinct values of~$\eta$ will be referred to as `waterbags' since this term conjures up the correct mental image: parcels of phase space of a certain density that can be distorted and moved, but not rarefied, compressed, or superimposed. The mean phase-space density is then
\begin{equation}
\label{Eqn:S2:E2}
\crl{f}(\v{p}) =  \int \dd{\eta} \eta P(\v{p},\eta).
\end{equation}
Here, we have applied the intuition that the steady-state distribution function~$P(\v{p},\eta)$ will be homogeneous in space (this contrasts with Lynden-Bell's original treatment, which focused on gravitationally bound, and therefore inhomogeneous, systems). Lynden-Bell's statistical mechanics amounts to positing that, before the onset of `true' collisions,~$P(\v{p},\eta)$~will maximise the Gibbs--Shannon entropy 
\begin{equation}
\label{Eqn:S2:E3}
S = -\iint\dd{\v{p}}\dd{\eta} P(\v{p},\eta)\ln P(\v{p},\eta).  
\end{equation}
Note that the integral in $\eta$ must run over all the possible values, including $\eta = 0$ (the empty waterbag).

Equipped with an entropy ripe for maximisation, we must decide upon a set of reasonable constraints under which to maximise it. Naturally, since $P(\v{p},\eta)$ is a probability-density function in $\eta$ at a given $\v{p}$, its integral in $\eta$ at any $\v{p}$ must equal unity:
\begin{equation}
\label{Eqn:S2:E4}
\int\dd{\eta}P(\v{p},\eta) = 1.
\end{equation} 
As well as this, we fix the energy density of the system:
\begin{equation}
\label{Eqn:S2:E5}
\iint \dd{\v{p}}\dd{\eta} \varepsilon(\v{p})\eta P(\v{p},\eta) = E = \mathrm{const},
\end{equation}
where $\varepsilon(\v{p})$ is the energy of a particle as a function of its momentum $\v{p}$. Note that, within this formalism, one could include the interaction energy of particles with fields (electromagnetic, gravitational,  etc.), so that in its most general form $\varepsilon$ would be a function of both position and momentum, which could need to be solved self-consistently with~$P$. Here, we will neglect this rich complexity, assuming instead that, in the relaxed state, the energy of the fields has decayed to a negligible fraction of the total energy and, in the process of decaying, has mediated the relaxation of the distribution function. 

Next, we enforce the conservation of the Casimir invariants (\ref{Eqn:S2:E1}) by requiring that the volume-integrated probability of each waterbag stays constant, viz., 
\begin{equation}
\label{Eqn:S2:E6}
\int \dd{\v{p}} P(\v{p},\eta) = \rho(\eta) = \mathrm{const}.
\end{equation}
The function $\rho(\eta)$ will be referred to as the `waterbag content' and is determined by initial conditions. The waterbag content of the initial condition can be read off by integrating over all portions of phase space where the initial exact phase-space density is equal to a particular value, viz., 
\begin{equation}
\label{Eqn:S2:E7}
\rho(\eta) = \frac{1}{V}\iint\dd{\v{r}}\dd{\v{p}}\delta\big(\eta - f(\v{r},\v{p},t = 0)\big) = -\frac{1}{V}\dev{\rmGamma}{\eta},
\end{equation}
where $V$ is the system's spatial volume. \textit{A priori}, in Lynden-Bell's statistical mechanics, the degree of universality of the equilibrium distribution is determined by~$\rho(\eta)$: all initial conditions with the same waterbag content and energy lead to the same equilibrium. 

Maximising the entropy (\ref{Eqn:S2:E3}) subject to the constraints~(\ref{Eqn:S2:E4}),~(\ref{Eqn:S2:E5}), and~(\ref{Eqn:S2:E6})\footnote{We note that, while we have endowed the invariants~(\ref{Eqn:S2:E4}),~(\ref{Eqn:S2:E5}), and~(\ref{Eqn:S2:E6}) with special significance as constraints, there may be situations where additional invariants are necessary. For instance, in strongly magnetised plasmas, relaxation may occur before the conservation of particles' magnetic moments are broken. In such cases, further invariants would be necessary and would alter the character of the solution (cf. \citealt{helander_2017}). Here we shall consider only systems where the fields driving the relaxation may be arbitrary, but the only quantities conserved on the relaxation timescale are~(\ref{Eqn:S2:E4}),~(\ref{Eqn:S2:E5}), and~(\ref{Eqn:S2:E6}).} is equivalent to maximising, unconditionally, the functional
\begin{multline}
\label{Eqn:S2:E8}
S[P(\v{p},\eta)] - \int\dd{\v{p}}\lambda(\v{p})\left[\int \dd{\eta}P(\v{p},\eta) - 1\right] - \beta \left[\iint \dd{\v{p}}\dd{\eta} \varepsilon(\v{p}) \eta P(\v{p},\eta) - E \right] \\+ \beta\int \dd{\eta}  \eta \mu(\eta)\left[ \int \dd{\v{p}} P(\v{p},\eta) - \rho(\eta)\right],
\end{multline}
where $\lambda(\v{p})$, $\beta$ and $-\beta \eta \mu(\eta)$ are Lagrange multipliers. By analogy with textbook statistical mechanics, we will sometimes refer to~$\mu(\eta)$ as the `chemical potential' (which it is, being the Lagrange multiplier that fixes the number of particles in waterbag~$\eta$). Doing so, we find the Lynden-Bell equilibria
\begin{equation}
\label{Eqn:S2:E9}
P(\v{p},\eta) = \frac{e^{-\beta \eta\left[\varepsilon(\v{p}) - \mu(\eta) \right]}}{\int \dd{\eta'}e^{-\beta \eta'\left[\varepsilon(\v{p})- \mu(\eta') \right]}},
\end{equation}
where $\lambda(\v{p})$ has been computed explicitly to arrange for the correct normalisation~(\ref{Eqn:S2:E4}), whereas~$\beta$ and~$\mu(\eta)$ must be chosen in such a way as to satisfy the constraints~(\ref{Eqn:S2:E5}) and~(\ref{Eqn:S2:E6}). We note that, despite~(\ref{Eqn:S2:E6}), the mean phase-space density~$\crl{f}$, given by~(\ref{Eqn:S2:E2}), will, in general, not have the same level sets~(\ref{Eqn:S2:E1}) as the exact one~$f$ (since $\crl{f}$ is an averaged quantity). The equilibria~(\ref{Eqn:S2:E9}) are both homogeneous and isotropic: an inevitable consequence of the system having no preferred position or direction.

The similarity between the Lynden-Bell equilibria~(\ref{Eqn:S2:E9}) and the Fermi--Dirac distribution is immediately apparent\footnote{Indeed, the Fermi--Dirac distribution can be thought of as the special case of a two-level-set system, which further reduces to the Maxwell--Boltzmann distribution when degeneracy is neglected (see, e.g., \citealt{Chavanis_2006b, Ewart_2022} for details).}. This should come as no surprise, because phase-volume conservation functions analogously to Pauli's exclusion principle: pieces of the same waterbag, or different waterbags, cannot cohabit in phase space. The equilibria, therefore, have degeneracy effects incorporated within them.

The prescription for computing Lynden-Bell equilibria is now clear: given an initial condition, with the initial energy density~$E$ and waterbag content~$\rho(\eta)$ (determined by~(\ref{Eqn:S2:E7})), solve two coupled integral equations~(\ref{Eqn:S2:E5}) and~(\ref{Eqn:S2:E6}) with~$P(\v{p},\eta)$ given by~(\ref{Eqn:S2:E9}), determine~$\beta$ and~$\mu(\eta)$, and substitute back into~(\ref{Eqn:S2:E9}). Before considering the numerical solutions of this problem in section~\ref{Section:Numerical result}, we will first seek to understand the system analytically.
\section{Theory: degenerate and non-degenerate equilibria}
\label{Degenerate and Non-degenerate equilibria}
\subsection{Lynden-Bell equilibria as excited Gardner distributions}
\label{subsection:Lynden-Bell equilibria as excited Gardner distribution functions}
\begin{figure}
\centering
\includegraphics[width=1.0\textwidth]{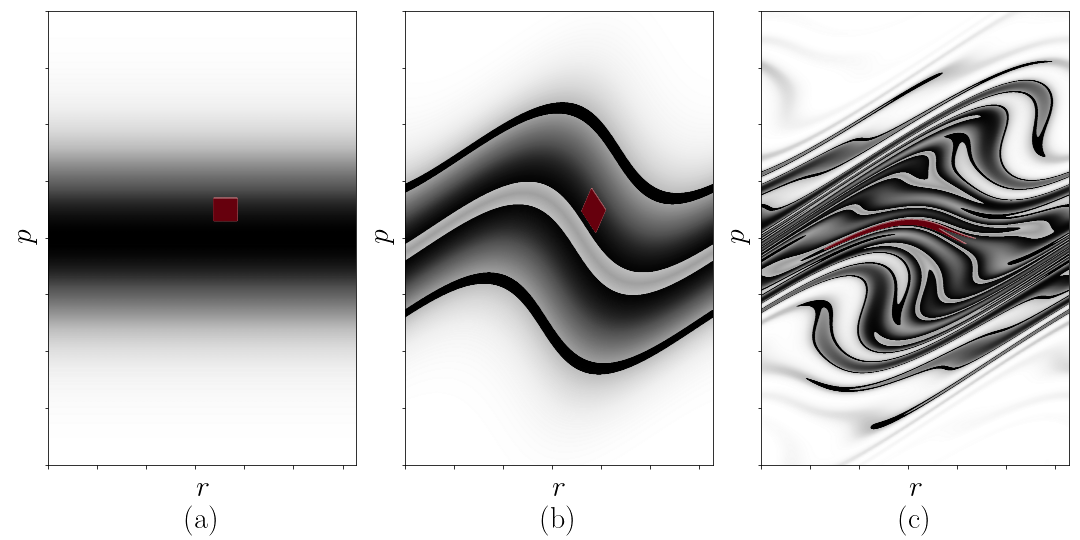}
\caption{A cartoon contour plot in phase space of three possible distribution functions, all of which possess identical waterbag contents. Panel (a) shows the Gardner distribution function corresponding to this waterbag content.  Panels (b) and (c) show distributions, at different (higher) energies, which can be reduced to the Gardner distribution by deforming and splicing the phase space incompressibly. A small patch of phase space is highlighted in red between plots to show the effect of the deformation.}
\label{Figure 1}
\end{figure}
Just as it is only meaningful to consider a Maxwellian with a positive energy, it is only meaningful to solve the Lynden-Bell equilibria (\ref{Eqn:S2:E9}) subject to reasonable choices of the constraints~(\ref{Eqn:S2:E5}) and~(\ref{Eqn:S2:E6}). It is therefore instructive to understand the properties of the waterbag-content function~(\ref{Eqn:S2:E6}). To get a feel for waterbag contents of typical initial conditions, one could compute the integral (\ref{Eqn:S2:E7}) for a range of examples (this is relatively simple due to the presence of the delta function). One quickly discovers that many different initial conditions have similar waterbag contents, just as many different initial conditions can have the same energy. To see this, we note that, from~(\ref{Eqn:S2:E7}), any volume-preserving transformation of the coordinates~$(\v{r},\v{p})$, including those transformations that `splice' the phase space discontinuously, will leave the waterbag content unchanged. This is unsurprising because the true, incompressible, flow of probability in phase space is precisely one such volume-preserving transformation. It is this freedom that implies that vastly different initial conditions can possess identical, or similar, waterbag contents. A cartoon illustrating this is given in Figure~\ref{Figure 1}, showing how seemingly complex distributions have the same waterbag content as very simple distributions. There will be families of initial conditions that have the same waterbag content, but different energies. 

For every family of initial conditions possessing the same waterbag content, there will be a unique distribution function that has that waterbag content but is a monotonically decreasing function of energy and, therefore, has the minimum possible energy associated with that waterbag content. Such a distribution function, for which the exact phase-space density satisfies~{$f(\v{r},\v{p}) = f_{\mathrm{G}}(\varepsilon(\v{p}))$}, is known as the Gardner distribution \citep{Gardner63,helander_2017}. The sequence of deformations of the distribution function to map an initial condition to its Gardner distribution function is often referred to as a `restacking', as it amounts to a reordering of phase-space elements into their minimum-energy configuration \citep{Dodin_2005, Kolmes_2020a, Kolmes_2020b}. Gardner distributions can be viewed as `ground states' associated with a given waterbag content (e.g., \citealt{helander_2017}), since no more energy can be extracted from such a distribution without violating phase-volume conservation. This fact intuitively guarantees that any initial condition that is a Gardner distribution is its own Lynden-Bell equilibrium since no other states are available to the system. Indeed, one can show that any Gardner distribution can be reconstructed from (\ref{Eqn:S2:E9}) for a particular choice of~$\beta$ and $\mu(\eta)$, although the proof is technical and left to Appendix \ref{Appendix: Degenerate Lynden-Bell equilibria}.

A generic initial condition can then be viewed as equivalent to taking some Gardner distribution and driving it out of equilibrium by the injection of some energy without changing the waterbag content. The Lynden-Bell equilibria are then simply the collisionless, phase-volume preserving, entropy-maximising equilibria of these higher-energy states, making them the natural excited states of Gardner distributions. Therefore, to capture the set of all possible waterbag contents, we need only study the set of all these `ground states', to which we would then add energy---the first step in the direction of universal outcomes.  Physically, this approach is equivalent to asking to what distribution a population of collisionless particles will relax once a certain amount of energy is injected into it---in a manner of speaking, a `thermodynamical' approach to `non-thermal' particle acceleration.

\subsection{Waterbag content of Gardner distributions}
\label{subsection:Waterbag content of Gardner distributions}
Having stated the problem in this way, we now consider the waterbag content associated with Gardner distributions. In what follows, we will consider Gardner distribution functions that are truncated at some minimum phase-space density~$\eta_{\mathrm{min}}$. This mathematical convenience will turn out to be a physical necessity. Thankfully, while it is mathematically and physically important that the cutoff~$\eta_{\mathrm{min}}$ be finite, it will only appear logarithmically in the outcomes of our calculations, making them highly insensitive to its actual value---yet another theory where the need for a cutoff is unavoidable but non-lethal.  

As a prototypical example, we compute $\rho(\eta)$ for a particular Gardner distribution: a truncated Maxwellian, viz.,
\begin{equation}
\label{Eqn:S3:E1}
f_{\mathrm{G}}(\v{p}) = \begin{cases}
\eta_{\mathrm{max}} e^{-\varepsilon(\v{p})/ \varepsilon_{0}} \quad & \text{for} \quad \varepsilon(\v{p}) < \varepsilon_{0}\ln\displaystyle\frac{\eta_{\mathrm{max}}}{\eta_{\mathrm{min}}}, \\[10pt] 0 \quad & \text{for} \quad \varepsilon(\v{p}) > \varepsilon_{0}\ln\displaystyle\frac{\eta_{\mathrm{max}}}{\eta_{\mathrm{min}}}. 
\end{cases}
\end{equation}
Besides $\eta_{\mathrm{min}}$, the parameters of this distribution are the energy scale~$\varepsilon_{0}$ and the maximum phase-space density~$\eta_{\mathrm{max}}$. The latter is straightforwardly related to the particle's spatial density, e.g.,~{${n_{0} = \eta_{\mathrm{max}}(2\pi m \varepsilon_{0})^{3/2}}$} in the limit~{${\eta_{\mathrm{min}}/ \eta_{\mathrm{max}} \to 0}$} for a 3D, non-relativistic plasma, where $\varepsilon(\v{p}) = p^{2}/2m$. The waterbag content of the Gardner distribution for such a plasma is then, from~(\ref{Eqn:S2:E7}),
\begin{equation}
\label{Eqn:S3:E2}
\rho(\eta) = \mathrm{\rmGamma}_{\mathrm{free}}\delta(\eta) + \begin{cases} \displaystyle\frac{2n_{0}}{\sqrt{\pi}\eta_{\mathrm{max}}\eta}\left(\ln\displaystyle\frac{\eta_{\mathrm{max}}}{\eta} \right)^{1/2} & \quad \text{for} \quad \eta_{\mathrm{min}} < \eta <\eta_{\mathrm{max}},
\\[10pt] 0 & \quad \text{otherwise},
\end{cases}
\end{equation}
where $\rmGamma_{\mathrm{free}}$ is the total volume of the momentum space that is unoccupied (i.e., where the exact phase-space density is zero). Of course, in reality, momentum space is unbounded and so $\rmGamma_{\mathrm{free}}$ is infinite. Formally, we are solving for the waterbag content and Lynden-Bell equilibrium in a momentum space of large, but finite, volume and will take this volume to infinity at the end of the calculation---of course, nothing physical will depend on $\rmGamma_{\mathrm{free}}$ as it becomes large. 

The presence of a $\delta(\eta)$ term in the expression for $\rho(\eta)$ is a generic feature, not restricted to the specific example (\ref{Eqn:S3:E2}). When it comes to solving (\ref{Eqn:S2:E6}) with $P(\v{p},\eta)$ given by (\ref{Eqn:S2:E9}), in order to find $\mu(\eta)$, this delta function can be accommodated by writing the chemical potential as 
\begin{equation}
\label{Eqn:S3:E3}
e^{\beta \eta \mu(\eta)} = \eta_{\mathrm{ref}}\delta(\eta) + \begin{cases}
\eta_{\mathrm{ref}}F(\eta) & \quad \text{for}\quad \eta_{\mathrm{min}} < \eta < \eta_{\mathrm{max}}, \\ 0 &\quad \text{otherwise},
\end{cases}
\end{equation}
where $\eta_{\mathrm{ref}}$ is some reference constant that must have dimensions of phase-space density. Its value is unimportant because $e^{\beta \eta \mu(\eta)}$ can always be rescaled by a constant without changing the Lynden-Bell equilibrium~(\ref{Eqn:S2:E9})---in essence,~$\eta_{\mathrm{ref}}$ is a gauge choice for the function $\mu(\eta)$. By analogy to textbook statistical mechanics, the function $F(\eta)$ will be referred to as the `fugacity' of the distribution. The form (\ref{Eqn:S3:E3}) results in the following expression for the Lynden-Bell equilibrium (\ref{Eqn:S2:E9}):
\begin{equation}
\label{Eqn:S3:E4}
P(\v{p},\eta) =  \frac{\delta(\eta) + e^{-\beta \eta \varepsilon(\v{p})}F(\eta)}{1 + \int_{\eta_{\mathrm{min}}}^{\eta_{\mathrm{max}}} \dd{\eta'}e^{-\beta \eta'\varepsilon(\v{p})}F(\eta')}.
\end{equation}
In (\ref{Eqn:S3:E4}), the first, $\delta(\eta)$, term in the numerator accounts for the probability density of finding phase space to be empty at a given location (this part of the phase space is referred to, aptly, as the `vacuum' by \citealt{Chavanis_2006b}), whereas the second term accounts for non-empty waterbags. Already~$\rmGamma_{\mathrm{free}}$ has dropped out of the calculation, as it must, and it is safe to let~$\rmGamma_{\mathrm{free}} \to \infty$. Likewise, it is immediately obvious that the reference phase-space density~$\eta_{\mathrm{ref}}$ has cancelled, as it also must. Thus, the Lynden-Bell equilibrium distribution~(\ref{Eqn:S3:E4}) depends only on the Lagrange multiplier~$\beta$ and the fugacity $F(\eta)$ (which themselves depend on~$\rho(\eta)$ for~$\eta > 0$ and the energy density $E$).
\begin{figure}
\centering
\includegraphics[width=1.0\textwidth]{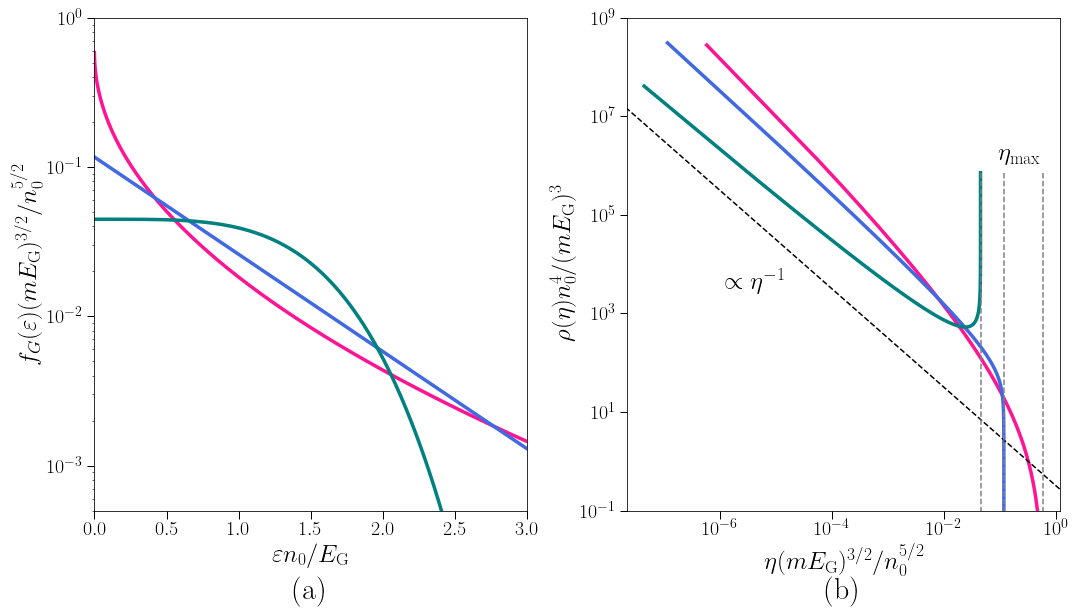}
\caption{(a) Three example Gardner distribution functions (the phase-space density here is plotted as a function of energy) and (b) their corresponding waterbag contents. All three distributions were chosen to have the same particle density~$n_{0}$ and energy density~$E_{\mathrm{G}}$. The maximum phase-space density~$\eta_{\mathrm{max}}$ of the distribution sets the upper cutoff of the waterbag content in~$\eta$, shown by the dashed vertical lines in (b). The lower cutoff~$\eta_{\mathrm{min}}$ is justified in Section \ref{Subsection:Minimum waterbag density}. We see that large differences at low~$\varepsilon$ only change the behaviour of $\rho(\eta)$ significantly at $\eta \sim \eta_{\mathrm{max}}$. For~$\eta\ll \eta_{\mathrm{max}}$, all three waterbag contents asymptote to a universal~$\eta^{-1}$ scaling.}
\label{Figure 2}
\end{figure}

The key feature of the example (\ref{Eqn:S3:E2}) is the~$\eta^{-1}$ power-law behaviour. While the specific form of the logarithmic factor in (\ref{Eqn:S3:E2}) was set by our (non-universal) choice of a Maxwellian Gardner distribution, the~$\eta^{-1}$ behaviour at small~$\eta$ is relatively universal. For any exponentially decaying Gardner distribution, i.e., for any distribution that at large energies can be approximated by~$\propto \exp\left[-\left(\varepsilon/\varepsilon_{0} \right)^{\sigma} \right]$, for some $\sigma > 0$ (or indeed can be bounded between two such functions), one will find a waterbag content with an~$\eta^{-1}$ asymptotic at~$\eta \ll \eta_{\mathrm{max}}$. 

To see this, we note that, since $f_{\mathrm{G}}(\varepsilon)$ is monotonically decreasing with energy, it has a well-defined inverse $f_{\mathrm{G}}^{-1}(\eta)$. In terms of this inverse, one can explicitly express the waterbag content (\ref{Eqn:S2:E7}) as 
\begin{equation}
\label{Eqn:S3:E5}
\rho(\eta) = \int\dd{\varepsilon}g(\varepsilon)\delta\left(\eta - f_{\mathrm{G}}(\varepsilon) \right) = -g\big(f_{\mathrm{G}}^{-1}(\eta)\big)\dev{f_{\mathrm{G}}^{-1}}{\eta}  \implies \dev{f_{\mathrm{G}}}{\varepsilon} = -\frac{g(\varepsilon)}{\rho\big(f_{\mathrm{G}}(\varepsilon)\big)},
\end{equation}
where $g(\varepsilon)$ is the density of states in energy, defined by the equation
\begin{equation}
\int \dd{\v{p}} (...) = \int \dd{\varepsilon}g(\varepsilon) (...).
\end{equation}
The first equality in (\ref{Eqn:S3:E5}) is a straightforward way to calculate the waterbag content of a given Gardner distribution, while the second is an equation from which the Gardner distribution can be constructed given knowledge of the system's waterbag content (cf. \citealt{Dodin_2005,helander_2017}). It is immediately clear why the~$\eta^{-1}$ scaling should arise in $\rho(\eta)$. For any exponentially decaying $f_{\mathrm{G}}(\varepsilon)$, the inverse function will be some logarithmic function of $\eta$, which, after differentiation in (\ref{Eqn:S3:E5}), will give an $\eta^{-1}$ asymptotic multiplied by some logarithmic function of $\eta$. To illustrate this, in Figure \ref{Figure 2}, we give three examples of starkly different Gardner distributions that, despite their differences, all possess waterbag contents which scale as~$\eta^{-1}$ at low~$\eta$. 

To convince a doubtful reader, we consider an alternative argument in support of the~$\eta^{-1}$ scaling of~$\rho(\eta)$. First, let us imagine a system in which we are allowed to vary~$\eta_{\mathrm{min}}$ freely while leaving the exact phase-space density~$f(\v{v})$ otherwise unchanged, as if there were some true distribution that had~$\eta_{\mathrm{min}} = 0$ and we were examining successive approximations to it (e.g., the difference between a truncated Maxwellian~(\ref{Eqn:S3:E1}) and a true Maxwellian). Let us consider the following integrals of such distribution functions (similar to the Casimir invariants considered by \citealt{Zhdankin_2021a}):
\begin{equation}
\label{Eqn:S3:E6}
\frac{1}{V}\iint_{f(\v{r},\v{p}) > \eta_{\mathrm{min}}}\dd{\v{r}}\dd{\v{p}} \left[f(\v{r},\v{p})\right]^{\gamma} = \int_{\eta_{\mathrm{min}}}^{\eta_{\mathrm{max}}} \dd{\eta} \eta^{\gamma}\rho(\eta) ,
\end{equation}where we have used~(\ref{Eqn:S2:E7}), and the phase-space integral is taken over the volumes where~{$\eta > \eta_{\mathrm{min}}$}. We have also used the property that the process of varying~$\eta_{\mathrm{min}}$ while otherwise leaving~$f(\v{r},\v{p})$ unchanged only changes the integration limits of~$\rho(\eta)$, without changing~$\rho(\eta)$ itself. The idea now is to vary the values of~$\eta_{\mathrm{min}}$ and~$\gamma$ and use what we know about~$f(\v{r},\v{p})$ to deduce the form of~$\rho(\eta)$. Clearly, for~$\gamma = 1$,~(\ref{Eqn:S3:E6}) is the particle density of the truncated~$f(\v{r},\v{p})$, which must be finite. This tells us that~$\rho(\eta)$ must integrate to a finite value when multiplied by~$\eta$. Furthermore, should~$f$ be any exponentially decaying function, then there would be a characteristic momentum scale above which the distribution is suppressed. This means that, as~$\eta_{\mathrm{min}}$ is taken to zero, both sides of (\ref{Eqn:S3:E6}) must converge for $\gamma = 1$. This is effectively a statement that the amount of probability contained beyond a few standard deviations is small. However, for an exponentially decaying phase-space density~$f$ and any positive power~$\gamma > 0$,~$f^{\gamma}$ is also exponentially decaying, so the same argument applies. Therefore, for exponentially decaying phase-space densities, $\rho(\eta)$ must be such that, when multiplied by any positive power of~$\eta$, it integrates to some finite value and is largely independent of the choice of lower cutoff~$\eta_{\mathrm{min}}$ of the integral. However, for~$\gamma = 0$, (\ref{Eqn:S3:E6}) becomes the (spatially averaged) momentum-space volume occupied by the truncated distribution: its support. For exponentially decaying distributions, which do not have compact support without truncation, this quantity will continue to grow without bound as $\eta_{\mathrm{min}}$ is decreased. Therefore,~$\rho(\eta)$ integrated with no powers of~$\eta$ must diverge as $\eta_{\mathrm{min}} \to 0$. It is obvious that~$\eta^{-1}$ is a function that has all these properties, but, more generally,~$\rho(\eta)$ could be any function of the form
\begin{equation}
\label{Eqn:S3:E7}
\rho(\eta) = \frac{n_{0}}{\eta_{\mathrm{max}}\eta}G\left(\frac{\eta}{\eta_{\mathrm{max}}} \right) \quad \text{for} \quad \eta_{\mathrm{min}} < \eta < \eta_{\mathrm{max}},
\end{equation}
where $G(x)$ is a dimensionless function whose dependence on~$x$ is weaker than any power law, viz.,
\begin{equation}
\label{Eqn:S3:E8}
\lim_{x \to 0} x^{\gamma - 1} G(x) = \begin{cases}
0 \quad &\text{if} \quad \gamma > 1, \\ \infty \quad &\text{if} \quad \gamma < 1. 
\end{cases}
\end{equation}
Note that the exact limit at $\gamma = 1$ cannot be determined by this argument; it is, in fact, dependent on the exact details of the exponential decay, but we will not require it for any calculations. This concludes the argument that Gardner distributions with exponential tails have waterbag contents with a universal low-$\eta$ asymptotic. 

It is a straightforward extension of this argument to work out what the waterbag content will be for Gardner distributions with non-exponential tails at high energies (low~$\eta$). Suppose that,  instead of being exponentially decaying, the Gardner distribution behaves as a power law at large momenta. Then there is a choice of~$\gamma > 0$ for which~$f^{\gamma}$, multiplied by the density of states in $|\v{p}|$, decays slower than~$|\v{p}|^{-1}$, implying that the integral~(\ref{Eqn:S3:E6}) will diverge as~$\eta_{\mathrm{min}}$ is decreased. This means that~$\rho(\eta)$ multiplied by~$\eta^{\gamma}$ for some $\gamma > 0$ must still have a divergent integral, so it must take the form
\begin{equation}
\label{eqn:S3:alphaweird}
\rho(\eta) = \frac{n_{0}}{\eta_{\mathrm{max}}^{2-\delta}\eta^{\delta}}G\left(\frac{\eta}{\eta_{\mathrm{max}}} \right) \quad \text{for} \quad \eta_{\mathrm{min}} < \eta < \eta_{\mathrm{max}},
\end{equation}
with some~$\delta > 1$. It turns out that~(\ref{eqn:S3:alphaweird}) also holds, but with~$\delta < 1$, for phase-space densities that go to zero algebraically even when~$\eta_{\mathrm{min}} = 0$\footnote{To apply this argument more generally to functions that have compact support in momentum space but do not go to zero algebraically~(e.g., step or bump functions), one can represent them as the limit of a sequence of functions that do go to zero algebraically.}. This is because the integral~(\ref{Eqn:S3:E6}) will be over a finite momentum-space volume even as~$\eta_{\mathrm{min}}$ is taken to zero, so, while~$\gamma < 0$ will make~$f^{\gamma}$ diverge is this finite domain, that divergence will still be integrable if~$\gamma$ is chosen sufficiently small and negative.

The explicit value of~$\delta$ can be found from~(\ref{Eqn:S3:E5}). To enable an explicit calculation, we will assume henceforth that the density of states~$g(\varepsilon)$ is related to energy by a simple power law, viz.,
\begin{equation}
\label{Eqn:S3:E11}
g(\varepsilon) = A\varepsilon^{a},
\end{equation}
where $A$ is an appropriate constant with dimensions~$[n_0/\eta_{\mathrm{max}}\varepsilon^{1+a}]$, and~$a$ is a real number. The assumption~(\ref{Eqn:S3:E11}) is not too restrictive as it can capture both non-relativistic and ultra-relativistic systems of any dimensionality. With the assumption~(\ref{Eqn:S3:E11}), we may now use~(\ref{Eqn:S3:E5}) to link the value of~$\delta$ to the high-energy asymptotic of the Gardner distribution. If the Gardner distribution has the power-law tail~$f_{\mathrm{G}}(\v{p}) \propto \varepsilon(\v{p})^{-(\chi + a)}$ at high energies, then, via~(\ref{Eqn:S3:E5}), one finds~{$\delta = 1 + (a+1)/(a+\chi)$}. If, instead, the Gardner distribution goes to zero at some finite energy~{$\varepsilon_{\mathrm{max}}$} so that~{$f_{\mathrm{G}}(\v{p}) \propto \left[\varepsilon_{\mathrm{max}} - \varepsilon(\v{p}) \right]^{\chi}$} near~$\varepsilon = \varepsilon_{\mathrm{max}}$, then~$\delta  = 1 - 1/\chi$.

Having catalogued the possible Gardner distributions and their corresponding $\rho(\eta)$, an important open question is now what Gardner distributions are the most common. Within the domain of numerical experiments, this is clearly decided by the whims of the numerical experimenter. However, it would seem reasonable to conjecture that in nature, the most common Gardner distributions should be ones with exponential tails. The reason for this is that the only processes that can change the Gardner distribution are, by definition, collisional, and collisional processes naturally relax the system to a Maxwell--Boltzmann distribution. More concretely, one often thinks of the sources of energy for violent relaxation as being large-scale inhomogeneities (e.g., counter-propagating flows, collapsing distributions of matter, etc.) of a system that is locally collisionally relaxed, and, therefore, has an exponential Gardner distribution.

We are now safe in the knowledge that the exponentially decaying Gardner distributions have~$\rho(\eta) \propto \eta^{-1}$, or, in more exotic cases,~$\rho(\eta) \propto \eta^{-\delta}$. We may now return to the task of solving (\ref{Eqn:S2:E5}) and (\ref{Eqn:S2:E6}) in light of these facts.   

\subsection{Non-degenerate Lynden-Bell equilibria}
\label{subsection:Non-degenerate Lynden-Bell equilibria}
Taking inspiration from the similarity between Fermi--Dirac and Lynden-Bell statistics, we may expect that there are two analytically tractable limits: degenerate (`cold') and non-degenerate (`hot'). In this section, we will explore the latter limit, which will turn out to be far more useful than the former (which is, nevertheless, also treated, for completeness, in Appendix \ref{Appendix: Degenerate Lynden-Bell equilibria}). 

We define the non-degenerate limit as one in which the probability of finding the exact phase-space density to be non-zero is small, viz.,
\begin{equation}
\label{Eqn:S3:E9}
D(\varepsilon) = \int_{\eta_{\mathrm{min}}}^{\eta_{\mathrm{max}}} \dd{\eta}e^{-\beta \eta\varepsilon}F(\eta) \ll 1.
\end{equation}
We shall call $D(\varepsilon)$ the `degeneracy parameter' since, from~(\ref{Eqn:S3:E4}), the probability that a position in phase space with a given energy is non-empty is given by the quotient~{$D(\varepsilon)/\left[1+D(\varepsilon)\right]$}. In the limit~(\ref{Eqn:S3:E9}), the distribution function~(\ref{Eqn:S3:E4}) can be approximated by
\begin{equation}
\label{Eqn:S3:E10}
P(\v{p},\eta) \simeq \delta(\eta)\left[1 - D\big(\varepsilon(\v{p})\big) \right] + e^{-\beta \eta\varepsilon(\v{p})}F(\eta).
\end{equation}
The effect of this simplification is  that the competition for any particular sub-volume of phase space is so weak that the waterbags are free to arrange themselves as Maxwellians~$\eta$ by~$\eta$. The waterbags with lower $\eta$ cost less energy to be placed at larger momenta---therefore, they have a larger thermal spread. In the non-relativistic limit, this is equivalent to the intuition that particles belonging to the waterbags with higher phase-space densities behave as though they have larger masses.

The approximate form (\ref{Eqn:S3:E10}) of the Lynden-Bell distribution makes computing the momentum-space integral in~(\ref{Eqn:S2:E6}) and determining the fugacity $F(\eta)$ a simple matter. Substituting~(\ref{Eqn:S3:E10}) into~(\ref{Eqn:S2:E6}) and using the explicit form~(\ref{Eqn:S3:E11}) of the density of states, we find the fugacity in the non-degenerate limit to be 
\begin{equation}
\label{Eqn:S3:E12}
F(\eta)  = \frac{(\beta \eta)^{1+a}}{A \Gamma(a+1)}\rho(\eta), \quad \eta_{\mathrm{min}} \leq \eta \leq \eta_{\mathrm{max}},
\end{equation}
where $\Gamma(a+1)$ is the gamma function. Substituting~(\ref{Eqn:S3:E12}) back into~(\ref{Eqn:S3:E10}) and using~(\ref{Eqn:S2:E2}) finally gives us an expression for the mean phase-space density (although still in terms of the as yet unspecified parameter $\beta$):
\begin{equation}
\label{eqn:S3:E10}
\crl{f}(\v{p}) =  \frac{\beta^{1+a}}{A \Gamma(a+1)}\int_{\eta_{\mathrm{min}}}^{\eta_{\mathrm{max}}}\dd{\eta} \eta^{2+a}\rho(\eta)e^{-\beta \eta \varepsilon(\v{p})}.
\end{equation}

From~(\ref{eqn:S3:E10}), it might seem as though, by diverse choices of~$\rho(\eta)$, a wide variety of distribution functions~$\crl{f}$ can be obtained. However, as we showed in section~\ref{subsection:Waterbag content of Gardner distributions}, a diversity of choices of waterbag contents is exactly what we do not have. Instead, fairly generic Gardner distributions with any form of exponential tails possess waterbag contents that are highly universal at low~$\eta$. Using~(\ref{Eqn:S3:E7}), we can make a convenient change of variables in~(\ref{eqn:S3:E10}),~$x = \beta\eta \varepsilon(\v{p})$, to find
\begin{equation}
\label{eqn:S3:E12}
\crl{f}(\v{p}) = \frac{n_{0}}{A\beta\Gamma(a+1) \eta_{\mathrm{max}}\varepsilon(\v{p})^{2+a}} \int_{\beta \eta_{\mathrm{min}}\varepsilon(\v{p})}^{\beta \eta_{\mathrm{max}}\varepsilon(\v{p})}\dd{x}x^{1+a}G\left(\frac{x}{\beta \eta_{\mathrm{max}}\varepsilon(\v{p})} \right)e^{-x}.
\end{equation}
This form exposes the fact that there is a natural power-law behaviour at energies such that 
\begin{equation}
\label{eqn:S3:E13}
\frac{1}{\beta \eta_{\mathrm{max}}} \ll \varepsilon(\v{p}) \ll \frac{1}{\beta \eta_{\mathrm{min}}}.
\end{equation}
This corresponds to the range of energies that are well within the thermal spread of the least dense waterbags, but far outside the thermal spread of the densest ones. At~{${\varepsilon(\v{p}) \gg 1/\beta \eta_{\mathrm{min}}}$}, the lower limit of the integral in~(\ref{eqn:S3:E12}) imposes an exponential cutoff on~$\crl{f}$. The distribution function $N(\varepsilon)$ of particle energies corresponding to (\ref{eqn:S3:E12}) is obtained by multiplying the mean phase-space density $\crl{f}(\v{p})$ by the density of states: 
\begin{equation}
\label{eqn:S3:E13p5}
N(\varepsilon) = g(\varepsilon)\crl{f}(\v{p}) = \frac{n_{0}}{ \beta \Gamma(a+1) \eta_{\mathrm{max}} \varepsilon^{2}}\int_{\beta \eta_{\mathrm{min}}\varepsilon}^{{\beta \eta_{\mathrm{max}}\varepsilon}}\dd{x} x^{1+a}G\left(\frac{x}{\beta \eta_{\mathrm{max}}\varepsilon} \right)e^{-x}.
\end{equation} 
This shows that the non-degenerate Lynden-Bell equilibria express a natural power-law tail and, furthermore, that this power law is independent of the type of plasma system under consideration. Note that the origin of the $\varepsilon^{-2}$ scaling found here is entirely different than the $\varepsilon^{-2}$ arising from particle acceleration in shocks \citep{Bell_1978}.

Consider now what happens if the Gardner distribution does not have an exponential tail. Then~$\rho(\eta)$ can be written as (\ref{eqn:S3:alphaweird}). Following all the same steps as before from~(\ref{eqn:S3:E10}) onwards, but using~(\ref{eqn:S3:alphaweird}) in place of~(\ref{Eqn:S3:E7}), one arrives at
\begin{equation}
\label{eqn:S3:E13p75}
N(\varepsilon) = \frac{n_{0}}{ \beta^{2-\delta} \Gamma(a+1) \eta_{\mathrm{max}}^{2-\delta} \varepsilon^{3-\delta}}\int_{\beta \eta_{\mathrm{min}}\varepsilon}^{{\beta \eta_{\mathrm{max}}\varepsilon}}\dd{x} x^{2+ a -\delta}G\left(\frac{x}{\beta \eta_{\mathrm{max}}\varepsilon} \right)e^{-x}.
\end{equation}
Thus, the resulting Lynden-Bell equilibrium again displays a power-law tail. The power law's exponent is set by the particular value of $\delta$ in (\ref{eqn:S3:alphaweird}), which is related to the Gardner distribution of that Lynden-Bell equilibrium via (\ref{Eqn:S3:E5}), as explained at the end of Section~\ref{subsection:Waterbag content of Gardner distributions}. For Gardner distributions that already have power-law tails, the resulting Lynden-Bell equilibria have shallower (ultra-`hard') power-law tails that strongly diverge in energy, giving the cutoff $\eta_{\mathrm{min}}$  pivotal importance. For Gardner distributions that decay faster than any exponential, the Lynden-Bell equilibria have `soft' power-law tails, with total energy depending only very weakly on the cutoff\footnote{The fact that certain choices of fugacity $F(\eta)$ give rise to Lynden-Bell equilibria that have power laws with various exponents was first noted, in connection to superstatistics, by \cite{Chavanis_2006b}.}.

Presently, however, we shall return to the Lynden-Bell equilibria~(\ref{eqn:S3:E13p5}) arising from exponential Gardner distributions, for which we complete the calculation.
\subsection{Calculation of $\beta$ and inevitability of partial degeneracy}
\label{subsection:The inevitability of partial degeneracy}
Both (\ref{eqn:S3:E13}) and (\ref{eqn:S3:E13p5}) still depend on the as yet unknown parameter~$\beta$, which, as well as fixing the energy, will determine the accuracy of the non-degeneracy approximation~(\ref{Eqn:S3:E9}). To find~$\beta$, we must compute the energy of our Lynden-Bell equilibrium~(\ref{Eqn:S3:E10}) according to~(\ref{Eqn:S2:E5}). Equivalently, from~(\ref{eqn:S3:E10}),
\begin{equation}
\label{eqn:S3:E14}
E  = \int\dd{\v{p}} \varepsilon(\v{p})\crl{f}(\v{p}) = \frac{a+1}{\beta}\int_{\eta_{\mathrm{min}}}^{\eta_{\mathrm{max}}} \dd{\eta} \rho(\eta).
\end{equation}
Therefore, in the non-degenerate limit,
\begin{equation}
\label{eqn:S3:E15}
\beta = \frac{a+1}{E}\int_{\eta_{\mathrm{min}}}^{\eta_{\mathrm{max}}} \dd{\eta}\rho(\eta).
\end{equation}
We see that $\beta$ decreases with increasing total energy of the distribution function; this is natural if $\beta$ is viewed as an inverse thermodynamic temperature. 

To see how this affects the underlying assumption~(\ref{Eqn:S3:E9}) of non-degeneracy, we evaluate~$D(\varepsilon)$ at~$\varepsilon \to 0$, where the degeneracy effect is strongest, since~$D(\varepsilon) \leq D(0)$. Requiring~$D(0) \ll 1$ gives the following condition on the total energy of the distribution, via~(\ref{Eqn:S3:E9}),~(\ref{Eqn:S3:E12}) and~(\ref{eqn:S3:E15}):
\begin{equation}
\label{eqn:S3:E16}
E \gg (a+1)\left[\int_{\eta_{\mathrm{min}}}^{\eta_{\mathrm{max}}} \dd{\eta} \frac{\eta^{a+1}\rho(\eta)}{A\Gamma(a+1)} \right]^{1/(a+1)}\int_{\eta_{\mathrm{min}}}^{\eta_{\mathrm{max}}} \dd{\eta}\rho(\eta).
\end{equation}
Since~$\rho(\eta)$ is a function only of the Gardner distribution, the right-hand side of~(\ref{eqn:S3:E16}) must scale with the energy density $E_{\mathrm{G}}$ of the Gardner distribution that has the same waterbag content~$\rho(\eta)$, but will always have~$E_{\mathrm{G}} \leq E$. For example, for the Gardner distribution (\ref{Eqn:S3:E1}), the right-hand side of (\ref{eqn:S3:E16}) should be proportional to $n_{0}\varepsilon_{0}$. Thus, just like in Fermi--Dirac statistics, the non-degeneracy approximation becomes more accurate as the distribution's energy density $E$ begins to dwarf the energy density $E_{\mathrm{G}}$ of the ground state. Note, however, that~(\ref{eqn:S3:E16}) contains an integral of the waterbag content $\rho(\eta)$ with no weighting by $\eta$, which, by~(\ref{Eqn:S2:E4}) and~(\ref{Eqn:S2:E6}), is the total phase volume occupied by non-empty waterbags.~Since~$\rho(\eta) \propto \eta^{-1}$ at small~$\eta$, this will be large, depending, albeit logarithmically, on the minimum waterbag density~$\eta_{\mathrm{min}}$. Indeed, for our example~(\ref{Eqn:S3:E1}),~(\ref{eqn:S3:E16}) becomes
\begin{equation}
\label{Eqn:S3:E22}
E \gg \frac{4}{3\sqrt{\pi}}n_{0}\varepsilon_{0}\left(\ln\frac{\eta_{\mathrm{max}}}{\eta_{\mathrm{min}}} \right)^{3/2}\left[ 1 + \mathcal{O}\left(\frac{\eta_{\mathrm{min}}}{\eta_{\mathrm{max}}} \right) \right]. 
\end{equation}
This means that the condition~(\ref{eqn:S3:E16}) requires the energy of the distribution to be much greater not just than the energy of the corresponding Gardner distribution, but than the Gardner energy multiplied by a polylogarithmic function of $\eta_{\mathrm{min}}$. This is a manifestation of the fact that, as $\eta_{\mathrm{min}}$ is taken to zero, more of the phase space is pervaded by low-density waterbags, making true non-degeneracy harder to achieve. It is thus impossible to achieve a non-degenerate limit unless $\eta_{\mathrm{min}}$ is kept finite. 

This is not the only place where the finiteness of the cutoff~$\eta_{\mathrm{min}}$ has raged against the dying of the light. The same effect is manifest in the power law of $\varepsilon^{-2}$ appearing in~(\ref{eqn:S3:E13p5}), which would have led to a logarithmically divergent mean particle energy were it not for the exponential cutoff at~$\varepsilon \sim 1/\beta \eta_{\mathrm{min}}$. This is obvious in~(\ref{eqn:S3:E14}), where the integral of~$\rho(\eta)$ has the same logarithmic divergence with~$\eta_{\mathrm{min}} \to 0$ as it did in the right-hand side of~(\ref{eqn:S3:E16}). This then makes its way into the expression (\ref{eqn:S3:E15}) for $\beta$, so, formally in the limit~$\eta_{\mathrm{min}} \to 0$,~$\beta \to \infty$ always! 

Since we must keep $\eta_{\mathrm{min}}$ finite, it is of substantial importance to understand the physical significance of it and the extent to which one should be prepared to accept one's equilibrium's dependence on its value.
\subsection{Physical meaning of minimum waterbag density}
\label{Subsection:Minimum waterbag density}
A Lynden-Bell equilibrium is essentially the thermal equilibrium of a collection of correlated blobs in phase space. This is to say that, inherent to the idea of computing the mean phase-space density, we have assumed that an exact phase-space density, i.e., a finite value of~$\eta$, is a meaningful concept. But, of course, an exact phase-space density is a fiction, since a plasma is composed of many discrete particles, and a phase-space density is only an average occupation number of particles' positions in phase space. The only sense in which an exact, continuous, phase-space density can be meaningful in a collisionless plasma then is if, within a small enough phase-space volume~$\delgam$, many particles can be considered to move as a collective entity: a waterbag. Then, on the scale of~$\delgam$, the system is composed of many waterbags with some `exact' phase-space density, whereas on scales much larger than~$\delgam$, the system can attain a mean phase-space density. This `correlation volume' provides a natural way to introduce the minimum non-zero phase-space density: clearly that should correspond to a single particle sitting in~$\delgam$, giving~{${\eta_{\mathrm{min}} = \delgam^{-1}}$}. 

Determining the value of~$\delgam$, is, however, a non-trivial challenge (see, e.g., discussions in \citealt{Kadomtsev_Pogutse70,Chavanis_2021,Ewart_2022}). \cite{Ewart_2022} argued, on the grounds that any meaningful collisionless-relaxation rate must be smaller than the plasma frequency but larger than the rate at which collisions break phase-volume conservation, that a reasonable constraint to place on the correlation volume is
\begin{equation}
\label{eqn:S3:E17}
\delgam \sim \frac{1}{\eta_{\mathrm{eff}}}\left(n_{0} \lambda_{D}^{3} \right)^{\alpha}, \quad \frac{2}{3} < \alpha < 1,
\end{equation}
where~$\eta_{\mathrm{eff}}$ is some typical phase-space density, which we can here estimate by~$\eta_{\mathrm{max}}$, and~$\lambda_{D}$ is the Debye length. This gives us an estimate for the minimum waterbag density: 
\begin{equation}
\label{eqn:S3:E18}
\eta_{\mathrm{min}} = \delgam^{-1} \sim \eta_{\mathrm{max}} \left(n_{0} \lambda_{D}^{3} \right)^{-\alpha}.
\end{equation}

Since the typical number of particles in a Debye sphere can be as large as~$10^{6}$ to~$10^{8}$ in collisionless  plasma environments (such as the solar wind or interstellar gas: see, e.g., \citealt{Verscharen_2019} and \citealt{Ferriere_2019}), the estimate (\ref{eqn:S3:E18}) might seem damningly small. It is, in fact, ideal. The existence of a broad power-law tail in~(\ref{eqn:S3:E13p5}) required a scale separation between~$\eta_{\mathrm{max}}$ and~$\eta_{\mathrm{min}}$. The estimate (\ref{eqn:S3:E18}) certainly provides this separation, tied to the plasma parameter. As for the breakdown of the non-degenerate approximation and the marginal divergence of the mean particle energy of a distribution with an~$\varepsilon^{-2}$ power law, we are saved by the fact that only the logarithm of the ratio~$\eta_{\mathrm{max}}/\eta_{\mathrm{min}}$ will appear, which, while large, can only ever be in the range of~{$10 - 30$}. This is somewhat reminiscent of the situations in the conventional theory of Coulomb collisions in plasmas, where forcible introduction of a phase-space cutoff into the collision integral results in only a weak dependence on the value of this cutoff, via the so-called Coulomb logarithm (see, e.g., \citealt{Helander_book}). 

Nevertheless, the presence of the logarithmic divergence with~$\eta_{\mathrm{min}}$ in the expression for~$\beta$, signposted at the end of section \ref{subsection:The inevitability of partial degeneracy}, will make it difficult to satisfy the non-degeneracy approximation\footnote{Although, notably, the fully non-degenerate limit can be naturally recovered by numerical noise in particle-in-cell (PIC) codes: see appendix \ref{Section:PIC}.}. There is good reason to suppose, however, that its breakdown may only be partial. We note that evaluating (\ref{Eqn:S3:E9}) at $\varepsilon = 0$ is tantamount to requesting non-degeneracy everywhere in the distribution. Since the degeneracy parameter $D(\varepsilon)$ decreases with increasing energy, it is reasonable to expect (and indeed we will see) some degeneracy at low energies, which gives way to non-degeneracy at higher energies, where our power-law tails will be recovered. As ever, the true solution lies on the cusp of asymptotic theory, and to go any further we must resort to numerical computation.
\section{Numerical results: partially degenerate equilibria}
\label{Section:Numerical result}
In this section, we shall recover the analytically  predicted power-law tail in (\ref{eqn:S3:E13p5}) by solving the constraint equations (\ref{Eqn:S2:E5}) and (\ref{Eqn:S2:E6}) for the Lynden-Bell equilibria (\ref{Eqn:S3:E4}). The numerical scheme for this is documented in detail in Appendix \ref{Section:Numerical Method}, amounting to an iterative method coupled to a~1D root finder. For these numerical results, we have restricted ourselves to a 3D, non-relativistic plasma with $\varepsilon(\v{p}) = p^{2}/2m$, although we anticipate from Section \ref{Degenerate and Non-degenerate equilibria} that these results extend, qualitatively, to general regimes. 

To capture a broad range of initial conditions, we consider a family of waterbag contents defined by
\begin{equation}
\label{eqn:S4:E1}
\rho(\eta) = \frac{4\pi p_{\mathrm{th,}\sigma}^{3}}{\sigma \eta}\left(\ln \frac{\eta_{\mathrm{max,}\sigma}}{\eta} \right)^{(3-\sigma)/\sigma}, \quad \eta_{\mathrm{min}} < \eta < \eta_{\mathrm{max,}\sigma},
\end{equation}
with $\sigma > 0$. This defines the family of Gardner distributions
\begin{equation}
\label{eqn:S4:E3}
f_{\mathrm{G,}\sigma}(\v{p}) = \begin{cases}
\eta_{\mathrm{max},\sigma}e^{-\left(p/p_{\mathrm{th,}\sigma}\right)^{\sigma}} \quad & \text{for} \quad p < p_{\mathrm{th,}\sigma}\left(\ln \displaystyle\frac{\eta_{\mathrm{max,}\sigma}}{\eta_{\mathrm{min}}} \right)^{1/\sigma}, \\[8pt] 0 \quad  & \text{for} \quad p > p_{\mathrm{th,}\sigma}\left(\ln \displaystyle\frac{\eta_{\mathrm{max,}\sigma}}{\eta_{\mathrm{min}}} \right)^{1/\sigma},
\end{cases}
\end{equation}
which, in the limit of~$\eta_{\mathrm{min}}/\eta_{\mathrm{max,}\sigma} \to 0$, have particle densities~$n_{0}$ and energy densities~$E_{\mathrm{G}}$ that satisfy
\begin{equation}
n_{0} = \frac{4\pi}{\sigma}\Gamma\left(\frac{3}{\sigma}\right)p_{\mathrm{th,}\sigma}^{3} \eta_{\mathrm{max,}\sigma}, \quad E_{\mathrm{G}} = \frac{\Gamma(5/\sigma)}{\Gamma(3/\sigma)}n_{0}\frac{p_{\mathrm{th,}\sigma}^{2}}{2m}.
\end{equation}
\begin{figure}
\centering
\includegraphics[width=1.0\textwidth]{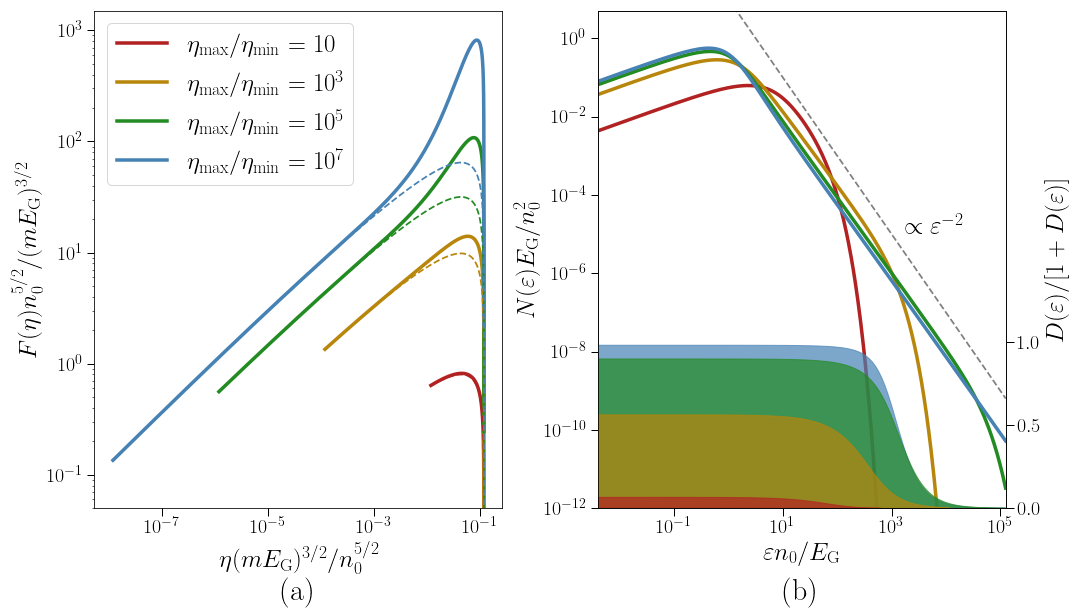}
\caption{Numerically computed Lynden-Bell equilibria for a range of~$\eta_{\mathrm{max,}\sigma}/\eta_{\mathrm{min}}$ and with~$\rho(\eta)$ given by~(\ref{eqn:S4:E1}) with~$\sigma = 2$. The energy density is equal to~$10 E_{\mathrm{G}}$ in all cases.~(a)~The numerically computed fugacity~$F(\eta)$ (solid lines) compared with the analytical solution~(\ref{Eqn:S3:E12}) obtained in the non-degenerate limit (dashed lines).~(b)~The resulting distributions $N(\varepsilon)$ of particle energies, with the universal power law~$\propto \varepsilon^{-2}$ shown for reference, cf.~(\ref{eqn:S3:E13p5}). Overplotted in solid colour (with the value range shown on the right) is the level of degeneracy~$D(\varepsilon)/\left[1+ D(\varepsilon)\right]$ (the probability that a given energy is occupied by a non-empty waterbag) as a function of energy;~$D(\varepsilon)$ is defined in~(\ref{Eqn:S3:E9}).}
\label{Figure 3}
\end{figure}Since these Gardner distributions represent minimum-energy states, we will be able to scan in the energy density of the system for all~$E \geq E_{\mathrm{G}}$, imagining some initial distribution of particles, with waterbag content $\rho(\eta)$ and energy density~$E_{\mathrm{G}}$, being accelerated to the energy density~$E$, and then seeking a maximum-entropy state (see Section~\ref{subsection:Lynden-Bell equilibria as excited Gardner distribution functions}). We can also vary~$\sigma$ in order to see the effects of the shape of the underlying Gardner distribution on the resulting Lynden-Bell equilibria.
\subsection{Degrees of degeneracy}
Let us first scan in~$\eta_{\mathrm{max,}\sigma}/\eta_{\mathrm{min}}$ in order to show that we can indeed recover the fully non-degenerate limit solved in Section \ref{Degenerate and Non-degenerate equilibria}. Figure \ref{Figure 3} shows the result of such a scan for~$\sigma = 2$ and~$E = 10E_{\mathrm{G}}$. By comparing the exact (numerically calculated) fugacity to the theoretical prediction (\ref{Eqn:S3:E12}) obtained in the absence of phase-space degeneracy, we see that the agreement is nearly perfect when~$\eta_{\mathrm{max}}$ is close to~$\eta_{\mathrm{min}}$, e.g., when~$\eta_{\mathrm{max}}/\eta_{\mathrm{min}} = 10$. This is as expected, since the non-degenerate limit is valid when~(\ref{eqn:S3:E16}) holds, which it does, as can be confirmed from the solid red colour in panel (b), showing the probability~$D(\varepsilon)/\left[1 + D(\varepsilon)\right]$ (with~$D(\varepsilon)$ defined in~(\ref{Eqn:S3:E9})) that a region of phase space is occupied. However, at~$\eta_{\mathrm{max}}/\eta_{\mathrm{min}} = 10$, there is an insufficient range of waterbag levels to achieve the scale separation~(\ref{eqn:S3:E13}) necessary to resolve a power-law tail in energies. To see a power-law tail, one must increase~$\eta_{\mathrm{max}}/\eta_{\mathrm{min}}$ to higher values, but this comes at the price of increasing the degeneracy of the system and, hence, undermining the asymptotic regime in which the tail was derived in the first place. For the values of~$\eta_{\mathrm{max}}/\eta_{\mathrm{min}}$ that we argued in Section~\ref{Subsection:Minimum waterbag density} to be realistic, the system becomes strongly degenerate at low energies, which can again be seen from the solid colours in panel~(b). In spite of this, at high energies, the degeneracy falls away, and, correspondingly,~$F(\eta)$ at low~$\eta$ still agrees well with the non-degenerate approximation~(\ref{Eqn:S3:E12}). All this conspires to ensure that, even formally outside the non-degenerate limit, the power-law tail~$N(\varepsilon) \propto \varepsilon^{-2}$ is still manifestly present.
\begin{figure}
\centering
\includegraphics[width=1.0\textwidth]{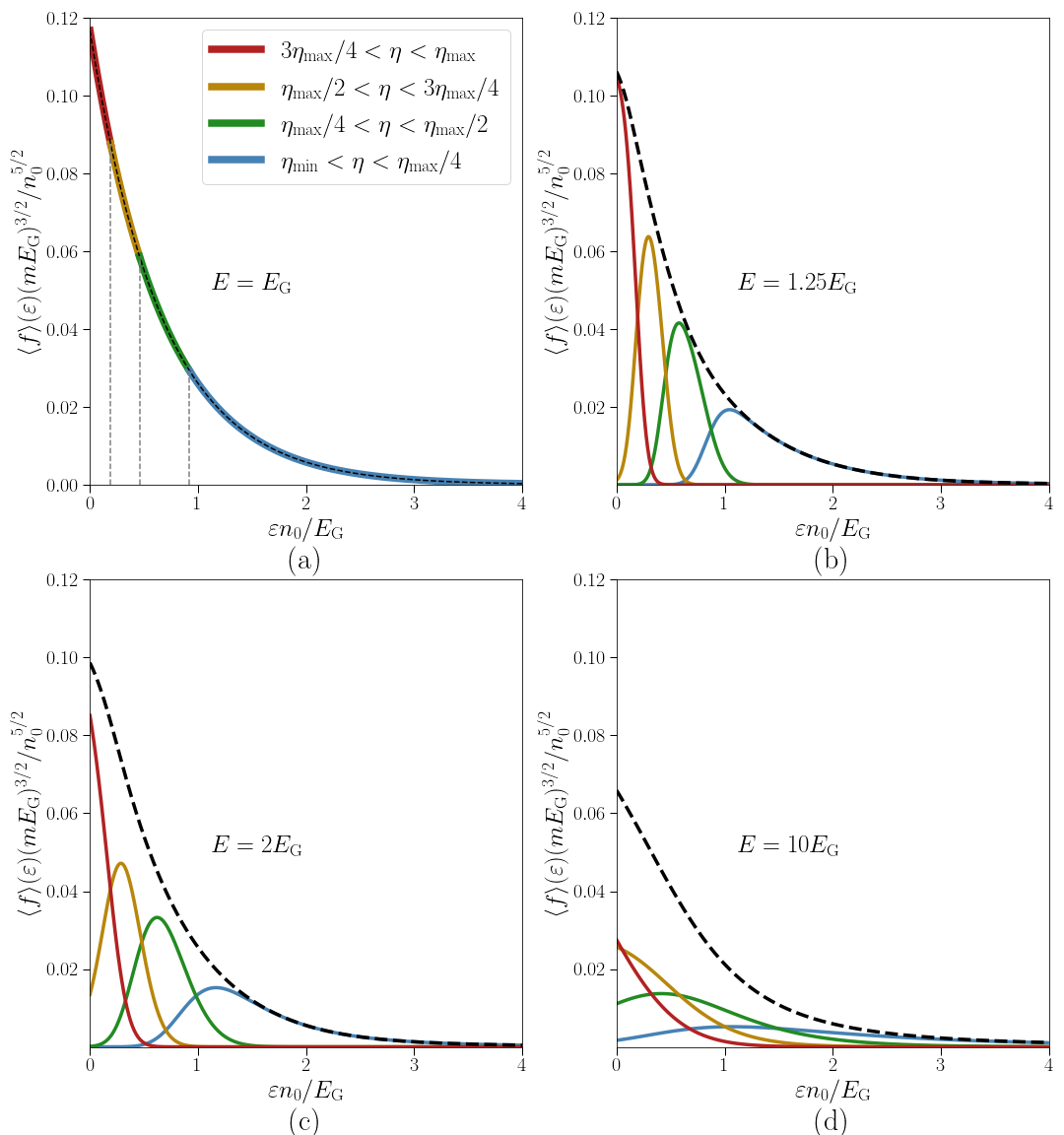}
\caption{Numerically computed Lynden-Bell equilibria for a range of energy densities~$E$ (in multiples of the energy density $E_{\mathrm{G}}$ of the underlying Gardner distribution) with~$\rho(\eta)$ given by~(\ref{eqn:S4:E1}) with~$\sigma = 2$ and~$\eta_{\mathrm{max}}/\eta_{\mathrm{min}} = 10^{6}$. In each plot, the dashed line is the mean phase-space density, while the underplotted solid lines are the contributions from four distinct ranges of exact phase-space density as functions of energy. Note that, while the exact phase-space densities have been grouped into four, this is not the same as solving for a four-waterbag Lynden-Bell equilibrium, as each grouping is still composed of a continuum of waterbags.}
\label{Figure split}
\end{figure}
\subsection{Energisation of particles and power-law tails}
\label{subsection:4.2}
\begin{figure}
\centering
\includegraphics[width=1.0\textwidth]{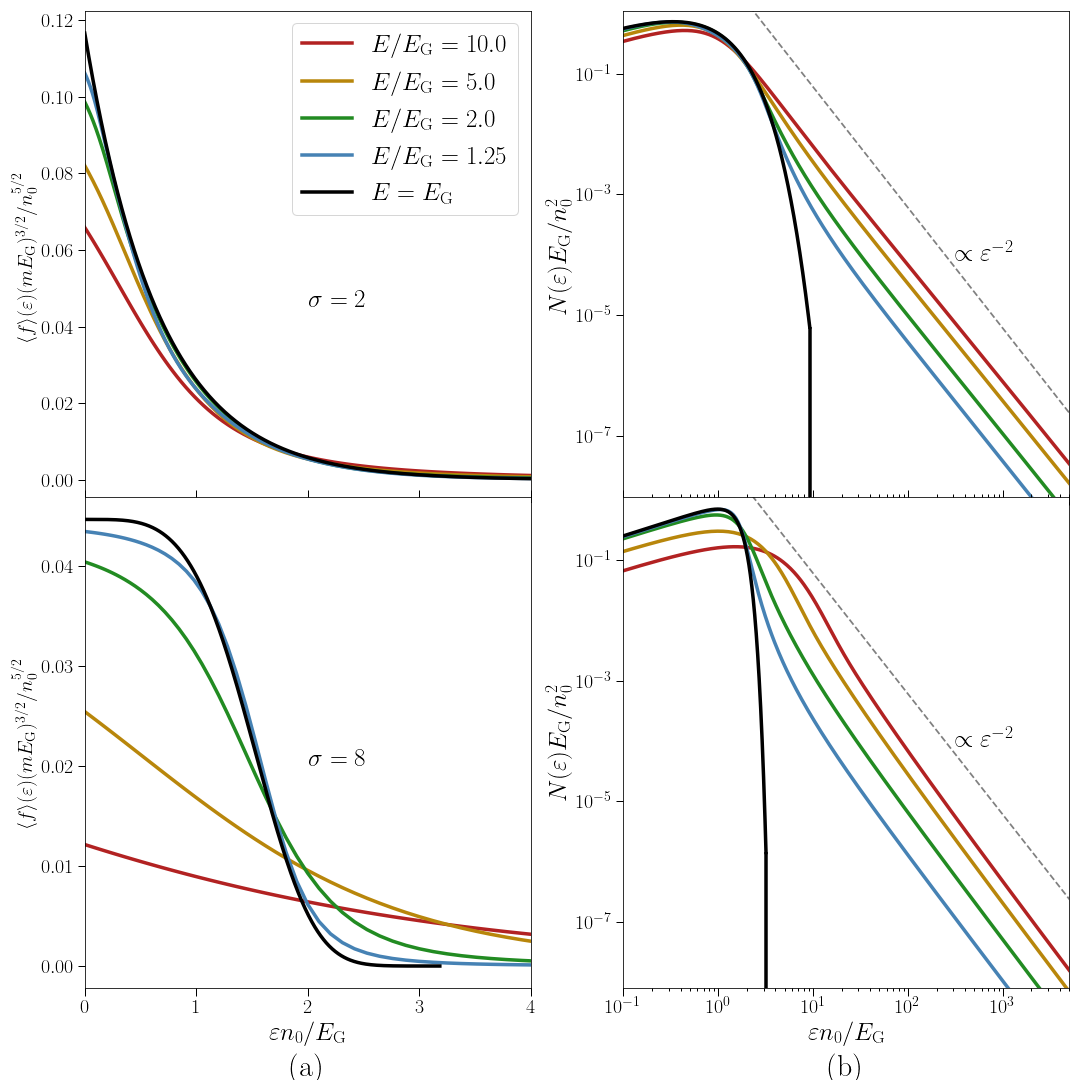}
\caption{Numerically computed Lynden-Bell equilibria with waterbag content given by the~$\sigma = 2$ (top) and~$\sigma = 8$~(bottom) cases of~(\ref{eqn:S4:E1}),~$\eta_{\mathrm{max,}\sigma} / \eta_{\mathrm{min}} = 10^{6}$.~(a) The phase-space densities shown in linear scale,~(b) the corresponding distributions of particle energies in logarithmic scale, for a range of ratios of~$E/E_{\mathrm{G}}$. The small deviations from the~$\varepsilon^{-2}$ tail can be attributed to the logarithmic corrections arising from the $x$ integral in (\ref{eqn:S3:E13p5}).}
\label{Figure 4}
\end{figure}Let us now scan in the energies densities $E$ of the distribution and again look for power-law tails and assess the effect of degeneracy. Figure \ref{Figure split} shows the results of such a scan, again {with~${\rho(\eta)}$} specified by~(\ref{eqn:S4:E1}) with~$\sigma = 2$. Plotted underneath the mean phase-space densities are the contributions from a number of finite ranges of exact phase-space density defined by
\begin{equation}
\crl{f}_{\eta_{1} < \eta < \eta_{2}}(\v{p}) = \int_{\eta_{1}}^{\eta_{2}} \dd{\eta} \eta P(\v{p},\eta).
\end{equation}
As one would anticipate, when $E$ is only slightly larger than $E_{\mathrm{G}}$, the effects of phase-space degeneracy are most prominent: the densest portions of the phase space clog up the lowest energies, forcing less dense portions to higher energies. As the total energy density is increased, we see that the contribution from each range of waterbags spreads out. This is because the increased energy allows dense portions of phase space to be promoted to larger energies, making room at lower energies for less dense portions of phase space to fill.

While the solutions plotted in Figure \ref{Figure split} might appear qualitatively similar to the Lynden-Bell equilibria obtained in numerical experiments with a small discrete number of level sets (see, e.g., \citealt{Assllani_2012} and \citealt{Ewart_2022}), this hides key universal features of systems with a continuum of level sets. To showcase this universality, in Figure~\ref{Figure 4}, we plot the Lynden-Bell equilibria for two different waterbag contents, $\sigma = 2$ and $\sigma = 8$ in~(\ref{eqn:S4:E1}), and a range of energy densities $E$. The $\varepsilon^{-2}$ power-law tails of these equilibria are immediately apparent, as predicted in (\ref{eqn:S3:E13p5}). 

Figure \ref{Figure 4} shows how these power-law tails become more prominent as one adds more energy to the Gardner distribution. At~$E = E_{\mathrm{G}}$, one has a highly non-universal Gardner equilibrium. As a small amount of energy~$E - E_{\mathrm{G}} \ll E_{\mathrm{G}}$ is added to~$E_{\mathrm{G}}$, the mean phase-space density at low energies is largely unaffected, while the power-law tail grows from the lowest-density waterbags. Thus, for energies close to the energy of the underlying Gardner distribution, the Lynden-Bell equilibria have a `core-halo' structure: the `core', which has energy density~$\sim E_{\mathrm{G}}$, is comprised of dense waterbags, which the system does not have sufficient energy to excite, whereas the `halo' is the tail comprised of those less dense waterbags that are capable of sampling a larger portion of phase space and thus arrange themselves into a universal~$\varepsilon^{-2}$ power law, containing the excess energy density~$\sim E - E_{\mathrm{G}}$. As the energy of the distribution is further increased, more waterbags have sufficient energy to sample a larger range of phase space, the halo continues to eat into the core, but both thermally broaden. At~$E \gg E_{\mathrm{G}}$, the asymptotically non-degenerate solution~(\ref{eqn:S3:E13p5}) with a power law in the energy range~(\ref{eqn:S3:E13}) suggests that the system strives for a state in which the halo has much more energy than the core. In this limit, one expects the transition between the core and halo to occur at~$\varepsilon \sim 1/\beta \eta_{\mathrm{max}}$. Since the halo should be exponentially suppressed at~$\varepsilon \gtrsim 1/\beta \eta_{\mathrm{min}}$, one can compute the ratio of the core energy to the halo energy. Owing to the~$\varepsilon^{-2}$ tail, this ratio will be proportional to~$\ln(\eta_{\mathrm{max}}/\eta_{\mathrm{min}})$ raised to some power, which depends on the specific functional form of the Gardner distribution. Whatever this power is, the vast majority of the total energy will be contained in the universal power-law tail for any system whose energy is much larger than that of its Gardner state.
\section{Conclusion}
\label{Section:Conclusion}
\subsection{Summary}
The \cite{LyndenBell67} equilibria are the natural maximum-entropy states for systems, such as a plasma described by the collisionless Vlasov equation~(\ref{Eqn:S1:E1}), which conserve not only density, momentum, and energy, but also an infinite family of further invariants~(\ref{Eqn:S2:E1}). These additional invariants are due to the conservation of phase volume, encoded by the `waterbag content' function~$\rho(\eta)$ given by (\ref{Eqn:S2:E7}) (equivalent to the Casimir invariants~(\ref{Eqn:S3:E6})), which measures the amount of phase volume where the exact phase-space density takes the value~$\eta$ (a `waterbag'), per unit~$\eta$. Maximising entropy subject to all these conservation laws then gives the mean phase-space density~(\ref{Eqn:S2:E2}) in the form of the Lynden-Bell equilibrium~(\ref{Eqn:S2:E9}) coupled with the constraints~(\ref{Eqn:S2:E5}) and~(\ref{Eqn:S2:E6}). In this paper, we have solved these constraint equations numerically in the general case, as well as analytically in a tractable limit (which turned out to be the practically relevant one). We have been able to show that, despite their apparent dependence on non-universal initial conditions, Lynden-Bell equilibria generically exhibit power-law tails, and, in particular, that a broad class of initial conditions will give rise to  the distribution of particles' energies scaling as $\varepsilon^{-2}$ at high $\varepsilon$.

To study the Lynden-Bell equilibria systematically, we first considered what values the invariants of the system, the energy density $E$ and waterbag content $\rho(\eta)$, could take. This led us to the concept of a Gardner distribution function, which is any monotonically decreasing function of the particle's energy. In Section \ref{subsection:Lynden-Bell equilibria as excited Gardner distribution functions}, we argued that to each possible initial condition one could assign a unique Gardner distribution, with the same waterbag content (and, therefore, the same Casimir invariants) as the initial condition, but a different energy, the Gardner distribution having, by definition, the lowest possible energy of all distributions with a given waterbag content $\rho(\eta)$. The Lynden-Bell equilibria at higher energies and the same $\rho(\eta)$ can, therefore, be thought of as excited states of this Gardner distribution. In Section \ref{subsection:Waterbag content of Gardner distributions}, we argued that the typical~$\rho(\eta)$ would have a fairly generic power-law form at low $\eta$, and, in particular, that it would scale as $\eta^{-1}$ for a wide class of initial conditions (see (\ref{Eqn:S3:E7})). In Section \ref{subsection:Non-degenerate Lynden-Bell equilibria}, we were able to find the Lynden-Bell equilibria analytically provided the energy of the system was sufficiently large for the competition of waterbags for phase space to be ignorable. In this `non-degenerate' limit, particles belonging to each waterbag arrange themselves into a separate Maxwellian equilibrium with an effective `temperature' inversely proportional to the phase-space density of that waterbag: denser portions of phase space are energetically costlier to move to higher energies. The resulting mean phase-space density~(\ref{eqn:S3:E10}) was found by integrating the contributions of all waterbags, each with their own Maxwellian distribution weighted by the amount of phase space which that waterbag occupied. Since the amount of each waterbag had a universal form~(\ref{Eqn:S3:E7}), this gave rise to a universal power-law tail~(\ref{eqn:S3:E13p5}) scaling as~$\varepsilon^{-2}$ at high particle energies. 

However, the non-degenerate limit required the system's energy to be asymptotically larger than the energy of the corresponding Gardner distribution. Formally, this turned out to be a very stringent limitation, and indeed one that could hardly ever be strictly fulfilled. Our analytical results were rescued by the argument, confirmed by the numerical solutions presented in Section \ref{Section:Numerical result}, that the effects of phase-space degeneracy were confined to the low-energy part of the distribution. The universal $\varepsilon^{-2}$ tail was numerically confirmed to be a robust feature of the generic Lynden-Bell equilibria. As well as ascertaining that a range of different initial conditions (\ref{eqn:S4:E1}) gave rise to the same power-law tail, the numerical solution also showed how this power-law tail formed. We found (Figure \ref{Figure 4}) that at energies comparable to the Gardner energy, the Lynden-Bell equilibria had a `core-halo' structure. The halo, consisting of the $\varepsilon^{-2}$ tail, was formed from low-density waterbags, which had sufficient energy to explore large portions of phase space, while the non-universal core was made up of denser waterbags, which did not have sufficient energies to be excited. As the energy of the Lynden-Bell equilibrium was increased, the halo ate its way into the core, making the distribution less and less degenerate and more universal in its shape. This behaviour is perhaps reminiscent of the measurements of the `non-thermal fraction' of particles in the solar wind (see, e.g., \citealt{Pierrard_2010,Oka_2015}).
\subsection{Limitations and applications}
The Lynden-Bell statistical mechanics presents an attractive scenario for the universal generation of power-law tails, which could perhaps offer insight into the distribution of particles in such astrophysical systems as cosmic rays and the solar wind. Indeed, a power law of $\varepsilon^{-2}$ in energy is roughly consistent with the observed power laws typical in the quiet-time solar wind (e.g., \citealt{Gloeckler2008,Fisk_2014,Yang_2020}) and close to, but distinct from, the inferred value for cosmic-ray sources (e.g., \citealt{Ormes_1978,Reichherzer_2021}). However, the universality of this predicted power law  fails to capture the wide range of power laws seen both in numerical simulations (e.g., \citealt{Sironi_2014,Zhdankin_2017,Werner_2017}) and observationally nearer to the Sun (see \citealt{Oka_2018} and references therein). Such systems are usually turbulent, possibly inhomogeneous, and invariably magnetised. In contrast, the theory that we have proposed for a universal power-law tail assumes a homogeneous system, in which all the fluctuating field's energy has decayed to a negligible fraction of the total energy. It is therefore an intriguing question whether our theory can be adjusted to apply to these cases. It is this question that we will address in this section, speculatively.

One common limitation in the applicability of Lynden-Bell's theory is that the fluctuating fields may decay away faster than the equilibrium state is reached---this is a well understood feature in galactic dynamics, where it is referred to as `incomplete relaxation' (see \citealt{Chavanis_2006a} and references therein). Here we have ignored such a possibility, assuming effectively that there will always be a sufficient amount of fluctuations to see the plasma through to its maximum-entropy state.

Perhaps an even more pressing concern is the possible reliance of the theory on precise phase-volume conservation. The validity of the Lynden-Bell statistics rests on the assumption that such an equilibrium can be reached long before the conservation of phase volume is broken. Conventional (linear) estimates for the timescale on which it would be broken, scaling as an inverse fractional power of the true Coulomb-collision frequency \citep{SuOberman1968}, indicate that such a relaxation should be possible. However, recent progress \citep{Beraldo_e_Silva_2017,Zhdankin_2021a, Nastac_2022} has shown that in turbulent systems, the conservation of phase volume may be broken on fast, collision-frequency-independent timescales. While it is possible that this is damning evidence against the existence (or persistence) of Lynden-Bell equilibria, it is also possible that it points to interesting interplay between the Lynden-Bell relaxation and collisional effects. For instance, one can imagine the possibility that the effect of collisions is to evolve~$\rho(\eta)$ without immediately pinning the system to a Maxwellian equilibrium. This can happen if collisions are already acting to erase small-scale phase-space structure of the exact phase-space density $f$ but not yet to change its mean $\crl{f}$ directly. In such a situation, one's aim would be to compute the evolution of $\rho(\eta)$ or, equivalently, of the underlying Gardner ground state $f_{\mathrm{G}}(\varepsilon)$. In particular, one can imagine a regime in which the underlying Gardner distribution evolves slower than the Lynden-Bell equilibrium is reached, causing the distribution to go through a sequence of Lynden-Bell equilibria. If the effect of collisions is to smooth the fluctuations $f - \crl{f}$  diffusively, it is clear that this can only cause the energy $E_{\mathrm{G}}$ of the Gardner distribution to increase or remain constant (cf. \citealt{Kolmes_2020b}). One would therefore expect that, as the system evolves through a sequence of Lynden-Bell equilibria, it will steadily become more degenerate as~$E_{\mathrm{G}}$ approaches the system's energy~$E$, with the core of the Lynden-Bell distribution eating into its tail (halo; see Section \ref{subsection:4.2}). This partially collisionless evolution would finally freeze once all fluctuations are diffused away, leaving a degenerate Lynden-Bell equilibrium doomed to further gradual erosion by weak Coulomb collisions on~$\crl{f}$. Since we have argued that an~$\varepsilon^{-2}$ tail is generic for any Lynden-Bell equilibrium whose underlying Gardner distribution has an exponential tail, this scenario suggests that the universal power-law tail could persist as long as $E_{\mathrm{G}} < E$, despite the breaking of phase-volume conservation due to weak collisionality. 

Finally, there is the question of whether any of this formalism can be ported smoothly to magnetised equilibria. Here, the most obvious straw to grasp at is that there is no guarantee that the invariants~(\ref{Eqn:S2:E5}) and (\ref{Eqn:S2:E6}) are the only invariants respected on the relaxation timescale. As previously mentioned, in drift-kinetic plasmas, each particle conserves its magnetic moment $\mu_{b}$. This implies a new conserved function
\begin{equation}
\label{Eqn:S5:E1}
\frac{2\pi}{V}\int\dd{\v{r}}\int\dd{v_{\parallel}}B(\v{r})\delta\big(\eta -  f(\v{r},v_{\parallel},\mu_{b})\big) = \rho(\eta, \mu_{b}),
\end{equation}
which would supersede the conservation of the now mundane $\rho(\eta)$. While some Gardner distributions have been studied for such systems (see \citealt{helander_2020,MackenBach_2022}), the Lynden-Bell equilibria in them are unexplored and may contain a wealth of interesting physics. This said, in turbulent systems, the conservation law~(\ref{Eqn:S5:E1}) may be just as fragile as the conservation of phase volume. Indeed it has been suggested that the breaking of adiabatic invariance may be essential to understanding the transport properties of non-thermal particles (see \citealt{ruszkowski2023cosmic} and references therein).

From the previous discussion it becomes clear that perhaps the most relevant limitation of Lynden-Bell's statistical mechanics---or indeed of any equilibrium statistical mechanics---in application to observed plasma phenomena is that much of real plasma dynamics are out of equilibrium in a physically essential way: any local relaxation processes, collisional or collisionless, tend to have to be taken into account alongside various `sources' and `sinks' of particles and/or energy, e.g., the energisation and escape of cosmic rays (\citealt{Schlickeiser_1989,Chandran_2000,Becker_Tjus_2020,Hopkins_2022,Kempski_2022}), the turbulent heating and radiative cooling of the intracluster medium (e.g., \citealt{Zhuravleva_2014} and references therein) or accretion-disc plasmas (e.g., \citealt{lesur_2021,kawazura_schekochihin_barnes_dorland_balbus_2022} and references therein), a veritable zoo of such processes in the solar wind (e.g., \citealt{Verscharen_2019,chen2020evolution}) and the Earth's magnetosphere (e.g., \citealt{lucek2005magnetosheath}), the birth of energetic $\alpha$-particles in fusion reactions and their subsequent slowing down and escape from confined plasmas (e.g., \citealt{Helander_book, Mailloux_2022}), etc. In plasmas where Coulomb collisions can be assumed to relax the particle distribution quickly to a local Maxwellian, we have a robust analytical framework for handling all these non-equilibrium processes in terms of the evolution of the density, momentum and temperature of that Maxwellian and a separation of the dynamics into that evolution plus the turbulence of small fluctuations around the local equilibrium (e.g., \citealt{Schekochihin_2009,Abel_2013}). In collisionless plasmas, such a framework is lacking as both the turbulence and the nature of the underlying equilibrium are mysterious and indeed it is not even guaranteed that they can be understood without detailed reference to each system's particular initial circumstances. If Lynden-Bell's statistical mechanics proves to be a viable collisionless substitute for Maxwell's, a path could be charted towards a theory of the dynamics and thermodynamics of collisionless plasmas possessing a modicum of universality. 

\section*{Acknowledgements}
We would like to thank Georgia Acton, Michael Barnes, Archie Bott, Andrew Brown, Steve Cowley, Jean-Baptiste Fouvry, Chris Hamilton, Per Helander, David Hosking, Matt Kunz, Ard Louis, Ralf Mackenbach, Ilya Nemenman, Sid Parameswaran, Patrick Reichherzer, Juan Ruiz Ruiz and Luis Silva  for illuminating discussions. The paper has also been improved by the recommendations of two anonymous reviewers. RJE’s work was supported by a UK EPSRC studentship. MLN was supported by a Clarendon Scholarship. The work of AAS was supported in part by grants from STFC (ST/W000903/1) and EPSRC (EP/R034737/1), as well as by the Simons Foundations via a Simons Investigator award.

Declaration of Interests: The authors report no conflicts of interest.
\appendix

\section{The degenerate limit of Lynden-Bell's statistics}
\label{Appendix: Degenerate Lynden-Bell equilibria}
In the main text, we have made use of the claim that, given the waterbag content $\rho(\eta)$, the Gardner distribution with the same waterbag content represents the ground state of all possible Lynden-Bell equilibria given by (\ref{Eqn:S3:E4}) that have this waterbag content. This is intuitive: should the initial condition of the system be a Gardner distribution, then that is, by definition, the only state available to the system, so it must also be the maximum-entropy state for that choice of $\rho(\eta)$. However, for completeness, and as a test of the Lynden-Bell formalism, it is prudent to check that the Gardner distribution can be recovered for some choice of the fugacity $F(\eta)$ and $\beta$. This is the aim of this appendix: to solve explicitly for $\beta$ and $F(\eta)$ in (\ref{Eqn:S3:E4}) when $E = E_{\mathrm{G}}$, where $E_{\mathrm{G}}$ is the energy of a given Gardner distribution $f_{\mathrm{G}}$, and to show that the mean phase-space density obtained by maximising Lynden-Bell's entropy is $\crl{f} = f_{\mathrm{G}}$.

To understand how the Gardner distribution will be recovered, we appeal to the familiar Fermi--Dirac distribution
\begin{equation}
f_{\mathrm{FD}}(\varepsilon) = \frac{\eta_{\mathrm{max}}e^{-\beta(\varepsilon - \mu)}}{1 + e^{-\beta(\varepsilon - \mu)}}.
\end{equation}
To work out what this is in the degenerate limit, every textbook notes that, when~$\beta$ is very large, the numerator (and second term in the denominator) is either very small for~$\varepsilon > \mu$, making the expression approximately zero, or very large for~$\varepsilon < \mu$, making the exponentials in the numerator and denominator approximately cancel to give $\eta_{\mathrm{max}}$. Borrowing this intuition, we anticipate that our solution should have~$\beta \to \infty$. It is clear what kind of solution one must search for: the degeneracy parameter~$D(\varepsilon)$ defined by (\ref{Eqn:S3:E9}) must be large wherever~$f_{\mathrm{G}}(\varepsilon)\neq 0$ and zero wherever~$f_{\mathrm{G}}(\varepsilon) = 0$. Furthermore, the dominant contribution to~$D(\varepsilon)$ in the integral~(\ref{Eqn:S3:E9}) must come from~$\eta = f_{\mathrm{G}}(\varepsilon)$. This is the Lynden-Bell version of the statement that the phase space is completely filled up below the Fermi energy. 

To see how this works in practice, let us posit the fugacity in the form  
\begin{equation}
\label{eqn:A1:E3}
F(\eta) = \frac{1}{\bar{\eta}}\exp\left[\beta\int_{\eta_{\mathrm{min}}}^{\eta}\dd{\eta'}f_{\mathrm{G}}^{-1}(\eta')\right]
\end{equation}
and prove that, via (\ref{Eqn:S3:E4}) and (\ref{Eqn:S2:E6}), it recovers the Gardner distribution with the correct waterbag content $\rho(\eta)$ when~$\beta \to \infty$, for a suitable choice of the dimensional constant~$\bar{\eta}$. 

In the denominator of~(\ref{Eqn:S3:E4}), we must evaluate the integral
\begin{equation}
\label{eqn:A1:E4}
D(\varepsilon) = \frac{1}{\bar{\eta}}\int_{\eta_{\mathrm{min}}}^{\eta_{\mathrm{max}}}\dd{\eta}e^{-\beta \eta\varepsilon + \ln \bar{\eta}F(\eta)}.
\end{equation}
Since we are working in the limit of large~$\beta$, this integral will be dominated by the contribution near the maximum (in~$\eta$ at fixed $\varepsilon$) of the exponent and thus can be evaluated by the method of steepest descent \citep{bender_book}. The location of the maximum of the exponent, which we denote by~$\eta_{\mathrm{stat}}(\varepsilon)$, is given by the solution to
\begin{equation}
\label{eqn:A1:E5}
\beta\dev{}{\eta}\left[\eta \varepsilon - \int_{\eta_{\mathrm{min}}}^{\eta}f_{\mathrm{G}}^{-1}(\eta') \dd{\eta'} \right] = 0 \implies \eta_{\mathrm{stat}}(\varepsilon) = f_{\mathrm{G}}(\varepsilon),
\end{equation}
as we anticipated above. We may now expand the exponent of the integrand in (\ref{eqn:A1:E5}) around this maximum to approximate the integral by
\begin{equation}
\label{eqn:A1:E6}
\begin{split}
D(\varepsilon)  & \simeq \int_{-\infty}^{\infty}\dd{\eta}\exp\left\lbrace-\beta f_{\mathrm{G}}(\varepsilon)\varepsilon+ \frac{\beta}{2}\left.\dev{f_{\mathrm{G}}^{-1}}{\eta}\right|_{\eta = f_{\mathrm{G}}(\varepsilon)}\left[\eta - f_{\mathrm{G}}(\varepsilon)\right]^{2}\right\rbrace F(f_{\mathrm{G}}(\varepsilon)) \\ & = e^{-\beta f_{\mathrm{G}}(\varepsilon)\varepsilon}F(f_{\mathrm{G}}(\varepsilon))\sqrt{\frac{2\pi}{\beta}}\left[-\left.\dev{f_{\mathrm{G}}^{-1}}{\eta}\right|_{\eta = f_{\mathrm{G}}(\varepsilon)} \right]^{-1/2}.
\end{split}
\end{equation}
As is customary, we have neglected the contributions from higher derivatives of the exponent in the knowledge that they will contribute terms that are smaller by $\mathcal{O}(1/\beta)$. We have also replaced the upper and lower limits of integration by~$\pm \infty$, assuming that the exponential decays sufficiently fast for the presence of integration limits not to be noticed by the integral. This will not be accurate near~$\varepsilon = 0$ and $\varepsilon = f_{\mathrm{G}}^{-1}(\eta_{\mathrm{min}})$, where the dominant contribution comes precisely from the limit of integration. This is, however, fine for~$\beta \to \infty$ because the intervals in~$\varepsilon$ where this approximation is bad shrink as~$\mathcal{O}\left(\beta^{-1/2}\right)$. 

If we further demand that the constant~$\bar{\eta}$ in (\ref{eqn:A1:E3}) is chosen so that
\begin{equation}
\frac{1}{\bar{\eta}}e^{-\beta \eta_{\mathrm{min}}f_{\mathrm{G}}^{-1}(\eta_{\mathrm{min}})}\sqrt{\frac{2\pi}{\beta}}\left[-\left.\dev{f_{\mathrm{G}}^{-1}}{\eta} \right|_{\eta = \eta_{\mathrm{min}}} \right]^{-1/2} = 1,
\end{equation}
then, neglecting terms of $\mathcal{O}(1/\beta)$,~$D(\varepsilon)$ satisfies
\begin{equation}
1+ D(\varepsilon) \simeq \begin{cases} D(\varepsilon) \quad & \text{for} \quad \varepsilon < f_{\mathrm{G}}^{-1}(\eta_{\mathrm{min}}), \\ 1 \quad & \text{for} \quad \varepsilon > f_{\mathrm{G}}^{-1}(\eta_{\mathrm{min}}).
\end{cases}
\end{equation}
Calculating the mean phase-space density~$\crl{f}(\v{p})$ from~(\ref{Eqn:S2:E2}) and (\ref{Eqn:S3:E4}), we find
\begin{equation}
\label{eqn:A2:E8}
\crl{f}(\varepsilon) \simeq \begin{cases} \displaystyle\frac{1}{\bar{\eta}D(\varepsilon(p))}\int_{\eta_{\mathrm{min}}}^{\eta_{\mathrm{max}}}\dd{\eta} \eta e^{-\beta \eta \varepsilon + \ln \bar{\eta}F(\eta)} \quad & \text{for} \quad \varepsilon < f_{\mathrm{G}}^{-1}(\eta_{\mathrm{min}}), \\[7pt] 0 \quad & \text{for} \quad \varepsilon > f_{\mathrm{G}}^{-1}(\eta_{\mathrm{min}}).
\end{cases}
\end{equation}

Let us prove that $\crl{f}(\v{p}) = f_{\mathrm{G}}(\varepsilon(\v{p}))$. The integral in (\ref{eqn:A2:E8}) can again be computed by the method of steepest descent. The exponent is the same as in (\ref{eqn:A1:E4}) and so will have the same maximum, at~$\eta_{\mathrm{stat}} = f_{\mathrm{G}}(\varepsilon)$ (up to a small~$\mathcal{O}(1/\beta)$ correction due to the factor of~$\eta$ in~(\ref{eqn:A2:E8})). Expanding the integral around this maximum and neglecting all terms that are~$\mathcal{O}(1/\beta)$ causes the factor of $\eta$ in the integral to be replaced by $\eta_{\mathrm{stat}}(\varepsilon)$. After this, the remainder of the integral has the same form as (\ref{eqn:A1:E6}), which cancels with the denominator of (\ref{eqn:A2:E8}), leaving only the factor of $\eta_{\mathrm{stat}} = f_{\mathrm{G}}(\varepsilon)$. Q.E.D.

Thus, the fugacity~(\ref{eqn:A1:E3}) correctly recovers the Gardner distribution~$f_{\mathrm{G}}(\varepsilon)$ in the limit of~$\beta \to \infty$. However, it is possible, in principle, to have accidentally chosen a fugacity which, while recovering the correct distribution, has an incorrect waterbag content. To complete the proof, we compute the waterbag content~(\ref{Eqn:S2:E6}) for the fugacity~(\ref{eqn:A1:E3}), to show that it is the same waterbag content as that of $f_{\mathrm{G}}$:
\begin{equation}
\begin{split}
\rho(\eta) & \simeq \int_{0}^{\infty}\dd{\varepsilon} g(\varepsilon) \exp\left\lbrace -\beta \varepsilon \left[\eta - f_{\mathrm{G}}(\varepsilon) \right] + \beta \int_{f_{\mathrm{G}}(\varepsilon)}^{\eta}f_{\mathrm{G}}^{-1}(\eta)\dd{\eta}\right\rbrace \\ & \quad \quad \quad \cdot\sqrt{\frac{\beta}{2\pi}}  \left[-\left.\dev{f_{\mathrm{G}}^{-1}}{\eta}\right|_{\eta = f_{\mathrm{G}}(\varepsilon)} \right]^{1/2}.
\end{split}
\end{equation}
Again using the method of steepest descent, we observe that the exponent has its stationary point, this time in~$\varepsilon$ at fixed~$\eta$, at~$\varepsilon =  f_{\mathrm{G}}^{-1}(\eta)$, just as we should expect. We can again expand the exponent around this stationary point and neglect $\mathcal{O}(1/\beta)$ terms, giving us
\begin{equation}
\begin{split}
\rho(\eta) \simeq & \int_{-\infty}^{\infty}\dd{\varepsilon}g(f_{\mathrm{G}}^{-1}(\eta))\exp\left\lbrace\frac{\beta}{2}\left.\dev{f_{\mathrm{G}}}{\varepsilon}\right|_{\varepsilon = f_{\mathrm{G}}^{-1}(\eta)} \left[\varepsilon - f_{\mathrm{G}}^{-1}(\eta)\right]^{2}\right\rbrace \sqrt{\frac{\beta}{2\pi}}\left[-\dev{f_{\mathrm{G}}^{-1}}{\eta} \right]^{1/2}  \\  = & -g\big(f_{\mathrm{G}}^{-1}(\eta)\big)\dev{f_{\mathrm{G}}^{-1}}{\eta}.
\end{split}
\end{equation}
This is exactly the expression (\ref{Eqn:S3:E5}) that we desire. This completes the proof that any Gardner distribution can be written as the $\beta \to \infty$ limit of the Lynden-Bell statistics, implying that one could, in principle, have discovered Gardner restacking just by analysing the Lynden-Bell equilibria at $\beta \to \infty$. 

By retaining terms that are small in~$\mathcal{O}(1/\beta)$, it is possible to analyse the Lynden-Bell equilibria analytically for energies very close but slightly above $E_{\mathrm{G}}$. However, in analogy to Fermi--Dirac statistics, this should just amount to broadening the `step' in the  Gardner distribution around~$\varepsilon = f_{\mathrm{G}}^{-1}(\eta_{\mathrm{min}})$, which limits the validity of this expansion to a very small range of energies, and thus makes such an expansion an exercise of infinitesimal utility.

\section{Numerical method for solving for Lynden-Bell equilibria}
\label{Section:Numerical Method}
In this appendix, we detail the numerical method by which we solve for the Lynden-Bell equilibria. To reiterate, the objective is, for a given waterbag content~$\rho(\eta)$ and energy density~$E$, to compute the function~$F(\eta)$ and the parameter~$\beta$ such that, with the waterbag distribution given by~(\ref{Eqn:S3:E4}), the constraints~(\ref{Eqn:S2:E5}) and~(\ref{Eqn:S2:E6}) are satisfied to sufficient accuracy. For the numerical solutions given in Section~\ref{Section:Numerical result}, the numerical method was tailored to a 3D, non-relativistic, system where $\varepsilon = p^{2}/2m$, but the method can be easily extended to any systems with a specified density of states.

The formula~(\ref{Eqn:S3:E4}) for the distribution function can be rewritten in such a way as to lend itself naturally to an iterative scheme. Namely, if one denotes the fugacity and the thermodynamic beta at the~$n^{\mathrm{th}}$ iteration by~$F^{(n)}$ and~$\beta^{(n)}$, respectively, then the natural iteration for the fugacity is
\begin{equation}
\label{Eqn:A1:E1}
F^{(n+1)}(\eta) = \rho(\eta)\left[2\pi\left(2m \right)^{3/2}\int_{0}^{\infty}\dd{\varepsilon} \varepsilon^{1/2} \frac{e^{-\beta^{(n)} \eta \varepsilon}}{1+ \int_{\eta_{\mathrm{min}}}^{\eta_{\mathrm{max}}}\dd{\eta'}e^{-\beta^{(n)} \eta' \varepsilon}F^{(n)}(\eta')}\right]^{-1}.
\end{equation}
Ignoring for a moment how one iterates~$\beta^{(n)}$, we note that if the iteration (\ref{Eqn:A1:E1}) converges, then by definition~(\ref{Eqn:S2:E6}) is satisfied. This means that a solution with the correct waterbag content has been found, but it may have an incorrect energy, since it does not necessarily satisfy~(\ref{Eqn:S2:E5}). However, by converging to a correct fugacity for a given~$\beta^{(n)}$, the problem is essentially reduced to a one-dimensional root-finding problem: finding~$\beta$ such that the energy takes the desired value, for which numerous numerical methods exist. This is the basis for our numerical algorithm, of which we will now give the specific details. 

The~$\eta$ domain is discretised into~$N_{\eta} = 10000$ points in preparation for future integration. To ensure that the lowest-density waterbags are well resolved without wasting resolution on the highest-density ones, a non-uniform discretisation in~$\eta$ is used: the~$j^{\mathrm{th}}$ grid point is given~by
\begin{equation}
\eta_{j} = \eta_{\mathrm{min}} + \left(\frac{j}{N_{\eta}} \right)^{q}\left(\eta_{\mathrm{max}} - \eta_{\mathrm{min}} \right),
\end{equation}
where the number~$q>1$ is chosen depending on the waterbag content: for (\ref{eqn:S4:E1}), $q = 3$ was used for~$\sigma \leq 3$ and~$q = 2$ for~$\sigma > 3$, to compromise on the resolution near~$\eta = \eta_{\mathrm{max}}$. 

An initial guess is chosen for the fugacity and thermodynamic beta, denoted~$F^{(0)}(\eta_{j})$ and~$\beta^{(0)}$, respectively. For the numerical solutions shown in Section~\ref{Section:Numerical result}, the initial guess was set to the analytical solutions (\ref{Eqn:S3:E12}) and (\ref{eqn:S3:E15}) obtained in the non-degenerate limit; other arbitrary choices were tested, all of which converged, albeit usually taking more iterations to do so. 

At each further iteration, the fugacity is updated according to~(\ref{Eqn:A1:E1}). The~$\eta'$ integral in~(\ref{Eqn:A1:E1}) is computed by a second-order midpoint method, interpolating the fugacity linearly between the neighbouring grid points. To compute the~$\varepsilon$ integral, a preliminary scan is first conducted to find the energy~$\varepsilon_{\mathrm{upper}}$ at which the degeneracy parameter (\ref{Eqn:S3:E9}) becomes smaller than~$10^{-5}$. This allows the energy integral in (\ref{Eqn:A1:E1}) to be split into two parts. The first of them, over energies below~$\varepsilon_{\mathrm{upper}}$, must be computed numerically, while the second, over energies above~$\varepsilon_{\mathrm{upper}}$, can be approximated by an analytically calculable function of~$\varepsilon_{\mathrm{upper}}$ and $\eta$. In the region where the integral must be computed numerically, the integration is carried out assuming the denominator to be piecewise linear on a momentum grid (rather than an energy grid, although this distinction is unimportant) linearly spaced with a spacing of~$10^{-3}$ in units such that the~$E_{\mathrm{G}} = 1$. The fugacity can thus be iterated at fixed thermodynamic beta. 

Once the integrated root-mean-square relative change in the fugacity over a single iteration is
\begin{equation}
\epsilon_{F} = \left[\frac{1}{\eta_{\mathrm{max}} - \eta_{\mathrm{min}}}\sum_{\eta_{j}}\left(\eta_{j+1} - \eta_{j} \right)\left(\frac{F^{(n+1)}}{F^{(n)}} - 1 \right)^{2} \right] ^{1/2} < 10^{-3},
\end{equation}
the energy of the resulting mean distribution can be computed. The root-finding method that we then use to determine~$\beta$ is an extremely primitive one: interval halving. The energy of the mean phase-space density with the fugacity resulting from above is computed, and~$\beta^{(n+1)}$ is then increased or decreased depending on whether the computed energy is too high or too low, respectively. The initial step size in $\beta$ is $\beta^{(0)}/2$. If the iteration in~$\beta$ passes over the root (i.e., in going from the~$n^{\mathrm{th}}$ iteration to the~$(n+1)^{\mathrm{st}}$, the energy goes from above the correct energy to below it, or vice versa), then the step size in~$\beta$ is halved, so that it eventually homes in on the root correctly. The step size is also halved if the step would otherwise result in a negative value for $\beta^{(n+1)}$. If the computed energy is within a tolerance of~$10^{-3}$ in units where~$E_{\mathrm{G}} = 1$, then the iteration in fugacity is allowed to proceed until~$\epsilon_{F}$ finally falls below~$10^{-5}$, at which point the solution is considered converged.

\section{Lynden-Bell equilibria and PIC plasmas}
\label{Section:PIC}
While we have shown numerically that the non-degenerate approximation of Lynden-Bell equilibria taken in section (\ref{subsection:Non-degenerate Lynden-Bell equilibria}) is qualitatively accurate even in systems that are nowhere near complete non-degeneracy, there is one (admittedly contrived) case where it is not just approximately true but represents an exact result. In this appendix, we show that non-degenerate Lynden-Bell equilibria are the natural long-time equilibria of a plasma which is evolved using the PIC algorithm (PIC plasma) in which any given true species is represented by PIC particles with multiple different weights. The intuitive reason for this is that PIC particles behave in a manner analogous to the `waterbags' central to the idea of Lynden-Bell relaxation. Waterbags are parcels of phase space, therefore containing some inherent number of true particles that move as a collective entity. PIC particles are hard-wired to represent such collections of true particles. To make this comparison more concrete, we will map a `collisionless collision operator', that describes the relaxation of a system to a Lynden-Bell equilibrium \citep{Ewart_2022} onto a numerical collision operator describing relaxation in a PIC plasma \citep{Touati_2022}---by mapping the physical picture of waterbags onto that of PIC particles.

The collisionless collision operator relaxes the probability density $P_{\alpha}(\v{v},\eta)$ of species~$\alpha$ as follows:
\begin{multline}
\label{eqn:A4:E17}
\pdev{P_{\alpha}}{t} = \sum_{\alpha'}\frac{q_{\alpha}^{2}q_{\alpha'}^{2}}{m_{\alpha}}\pdev{}{\v{v}}\cdot\int\dd{\v{v}'}{\sf{Q}}(\v{v},\v{v}')\cdot\int\dd{\eta'}\eta'\\ \Bigg\lbrace \frac{\delgam_{\alpha'}}{m_{\alpha}}\Big[ \eta' - f_{\alpha'}(\v{v}') \Big]P_{\alpha'}(\v{v}',\eta')\left.\pdev{P_{\alpha}}{\v{v}}\right|_{\eta} - \frac{\delgam_{\alpha}}{m_{\alpha'}}\Big[ \eta - f_{\alpha}(\v{v}) \Big]P_{\alpha}(\v{v},\eta)\left.\pdev{P_{\alpha'}}{\v{v}'}\right|_{\eta'}\Bigg\rbrace,
\end{multline}
where~$\delgam_{\alpha}$ is the typical volume over which a fluctuation in phase space is correlated (for details, see the discussion in section \ref{Subsection:Minimum waterbag density} or in \citealt{Ewart_2022}) and $\sf{Q}(\v{v},\v{v}')$ is a tensor containing information about the interaction potential, whose explicit form we will not need here. The derivation of collision operators such as (\ref{eqn:A4:E17}) is, naturally, subject to a number of approximations and caveats. Chief amongst these is the assumption of an electrostatic, quasilinear system in which phase volume is conserved. A full derivation and discussion of such collision operators can be found in, e.g., \cite{Chavanis_2021} or \cite{Ewart_2022}.

In a PIC simulation, a given true species of particle may be represented by a number of different macroparticles that have different `weights'---what this means quantitatively in our language, we shall explain shortly. To describe such a system, we set the distribution~$P_{\alpha}$ to be discrete in $\eta$, the latter taking values $\eta_{\alpha,a}$ corresponding to macroparticle `species' $a$:
\begin{equation}
P_{\alpha}(\v{v},\eta) = \sum_{a}P_{\alpha, a}(\v{v})\delta(\eta - \eta_{\alpha,a}).
\end{equation}
The collision operator (\ref{eqn:A4:E17}) then becomes
\begin{multline}
\label{eqn:A4:E18}
\pdev{P_{\alpha, a}}{t} = \sum_{\alpha'}\frac{q_{\alpha}^{2}q_{\alpha'}^{2}}{m_{\alpha}}\pdev{}{\v{v}}\cdot\int\dd{\v{v}'}{\sf{Q}}(\v{v},\v{v}')\,\cdot\sum_{a'}\\ \Bigg\lbrace \frac{\delgam_{\alpha'}\eta_{\alpha',a'}}{m_{\alpha}}\Big[ \eta_{\alpha',a'} - f_{\alpha'}(\v{v}') \Big]P_{\alpha',a'}(\v{v}')\pdev{P_{\alpha,a}}{\v{v}} - \frac{\delgam_{\alpha}\eta_{\alpha',a'}}{m_{\alpha'}}\Big[ \eta_{\alpha,a} - f_{\alpha}(\v{v}) \Big]P_{\alpha,a}(\v{v})\pdev{P_{\alpha',a'}}{\v{v}'}\Bigg\rbrace.
\end{multline}
Here, to reiterate,~$f_{\alpha}(\v{v})$ is the mean phase-space density of particles of species~$\alpha$, which can be written as the sum of the mean phase-space densities~$f_{\alpha, a}(\v{v})$ of different macroparticle `species':
\begin{equation}
\label{eqn:A4:E19}
f_{\alpha}(\v{v}) = \sum_{a}f_{\alpha,a}(\v{v}) = \sum_{a} \eta_{\alpha, a}P_{\alpha, a}(\v{v}).
\end{equation}
Next, one must note that PIC particles, like classical particles, occupy zero phase volume. Therefore, if all PIC particles are assumed decorrelated,~$\delgam_{\alpha}=0$. However, the phase-space densities~$\eta_{\alpha,a}$ are then infinite. Mathematically, this corresponds to taking the limit of~$\eta_{\alpha,a} \to \infty$ and~$\delgam_{\alpha} \to 0$ in~(\ref{eqn:A4:E18}) and~(\ref{eqn:A4:E19}) while holding the product~$\delgam_{\alpha}\eta_{\alpha,a}$---the number of particles in a correlated volume---fixed to the number of `true' particles~$\delta N_{\alpha,a}$ contained in a macroparticle of species~$(\alpha, a)$; the quantity~$\delta N_{\alpha,a}$ is what is usually called the `weight' of the macroparticle in the PIC terminology.

Clearly, as~$\eta_{\alpha, a}$ is taken to infinity, it is the mean phase-space density~$f_{\alpha,a} = \eta_{\alpha,a}P_{\alpha,a}$ that remains finite. Making all these substitutions and taking the appropriate limit, one finds from~(\ref{eqn:A4:E18}) that it relaxes according to
\begin{equation}
\label{eqn:A4:E20}
\pdev{f_{\alpha,a}}{t} = \sum_{\alpha', a'} \frac{q_{\alpha}^{2}q_{\alpha'}^{2}}{m_{\alpha}}\pdev{}{\v{v}}\cdot \int \dd{\v{v}'}{\sf{Q}} (\v{v},\v{v}')\cdot \left(\frac{\delta N_{\alpha' a'}}{m_{\alpha}}f_{\alpha', a'} \pdev{f_{\alpha, a}}{\v{v}} - \frac{\delta N_{\alpha, a}}{m_{\alpha'}}f_{\alpha, a}\pdev{f_{\alpha',a'}}{\v{v}'} \right).
\end{equation}
Modulo the details of the tensor~$\sf{Q}$ (due to the discrete nature of PIC codes), this is the effective collision operator due to numerical noise inherent in the PIC algorithm \citep{Boris_conferenceproc,birdsall1985plasma,Touati_2022}. It is easy to show that, provided the tensor~$\sf{Q}(\v{v},\v{v}')$ is positive definite and symmetric in~$(\v{v},\v{v}')$, the collision operator~(\ref{eqn:A4:E20}) has an H-theorem with the entropy
\begin{equation}
S = -\sum_{\alpha,a}\frac{1}{\delta N_{\alpha, a}} \int \dd{\v{v}} f_{\alpha,a}(\v{v})\ln f_{\alpha,a}(\v{v}),
\end{equation}
maximised by the equilibria 
\begin{equation}
f_{\alpha,a} = N_{\alpha, a}e^{-\beta \delta N_{\alpha,a}\varepsilon(\v{v})},
\end{equation}
where~$N_{\alpha,a}$ is a normalisation constant set by the number of macroparticles of each weight (the fugacity of these marcoparticles). The resulting distribution function of particles of true species~$\alpha$ is simply a superposition of Maxwellians:
\begin{equation}
f_{\alpha}(\v{v}) = \sum_{a}f_{\alpha, a}(\v{v}) = \sum_{a}N_{\alpha,a}e^{-\beta \delta N_{\alpha,a}\varepsilon(\v{v})},
\end{equation}
which is manifestly the discrete form of the non-degenerate Lynden-Bell {equilibrium~(\ref{eqn:S3:E10})}.

Thus, non-degenerate Lynden-Bell equilibria could emerge organically in PIC simulations where multiple macroparticle weights represent the same true particle species. Of course, this is more a numerical artefact than a physical result. The equilibrium towards which such a system is pushed by the numerical noise is effectively hard-coded by the choice of macroparticle weights. We note finally that the effects of a numerical collision operator such as~(\ref{eqn:A4:E20}) actually extend further than spurious consequences for the steady state. It was shown by \cite{Ewart_2022} that such collision operators could give rise to an anomalous interspecies drag, which again here would be a defect of the numerical method (cf. \citealt{May_2014}).

\bibliographystyle{jpp}

\bibliography{PLTbib}

\begin{thebibliography}{91}
\expandafter\ifx\csname natexlab\endcsname\relax\def\natexlab#1{#1}\fi
\def\au#1{#1} \def\ed#1{#1} \def\yr#1{#1}\def\at#1{#1}\def\jt#1{\textit{#1}}
  \def\bt#1{#1}\def\bvol#1{\textbf{#1}} \def\vol#1{#1} \def\pg#1{#1}
  \def\publ#1{#1}\def\arxiv#1{#1}\def\org#1{#1}\def\st#1{\textit{#1}}

\bibitem[Abel {\em et~al.\/}(2013)Abel, Plunk, Wang, Barnes, Cowley, Dorland \&
  Schekochihin]{Abel_2013}
{\sc \au{Abel, I.~G.}, \au{Plunk, G.~G.}, \au{Wang, E.}, \au{Barnes, M.},
  \au{Cowley, S.~C.}, \au{Dorland, W.} \& \au{Schekochihin, A.~A.}} \yr{2013}
  \at{Multiscale gyrokinetics for rotating tokamak plasmas: fluctuations,
  transport and energy flows}.  \jt{Rep. Prog. Phys.}  \bvol{76},  \pg{116201}.

\bibitem[Aitchison {\em et~al.\/}(2016)Aitchison, Corradi \&
  Latham]{Aitchison_2016}
{\sc \au{Aitchison, L.}, \au{Corradi, N.} \& \au{Latham, P.~E.}} \yr{2016}
  \at{Zipf’s law arises naturally when there are underlying, unobserved
  variables}.  \jt{PLoS Comput. Bio.}  \bvol{12},  \pg{1}.

\bibitem[Amato \& Casanova(2021)]{amato_casanova_2021}
{\sc \au{Amato, E.} \& \au{Casanova, S.}} \yr{2021}  \at{{On particle
  acceleration and transport in plasmas in the galaxy: theory and
  observations}}.  \jt{J. Plasma Phys.}  \bvol{87},  \pg{845870101}.

\bibitem[Arad \& Johansson(2005)]{Arad_2005}
{\sc \au{Arad, I.} \& \au{Johansson, P.~H.}} \yr{2005}  \at{A numerical
  comparison of theories of violent relaxation}.  \jt{Mon. Not. R. Astron.
  Soc.}  \bvol{362},  \pg{252}.

\bibitem[Assllani {\em et~al.\/}(2012)Assllani, Fanelli, Turchi, Carletti \&
  Leoncini]{Assllani_2012}
{\sc \au{Assllani, M.}, \au{Fanelli, D.}, \au{Turchi, A.}, \au{Carletti, T.} \&
  \au{Leoncini, X.}} \yr{2012}  \at{Statistical theory of quasistationary
  states beyond the single water-bag case study}.  \jt{Phys. Rev. E}
  \bvol{85},  \pg{021148}.

\bibitem[Balescu(1960)]{Balescu60}
{\sc \au{Balescu, R.}} \yr{1960}  \at{Irreversible processes in ionized gases}.
   \jt{Phys. Fluids}  \bvol{3},  \pg{52}.

\bibitem[Beck \& Cohen(2003)]{Beck_2003}
{\sc \au{Beck, C.} \& \au{Cohen, E.}} \yr{2003}  \at{Superstatistics}.
  \jt{Physica A}  \bvol{322},  \pg{267}.

\bibitem[{Becker Tjus} \& Merten(2020)]{Becker_Tjus_2020}
{\sc \au{{Becker Tjus}, J.} \& \au{Merten, L.}} \yr{2020}  \at{Closing in on
  the origin of galactic cosmic rays using multimessenger information}.
  \jt{Phys. Rep.}  \bvol{872},  \pg{1}.

\bibitem[{Bell}(1978)]{Bell_1978}
{\sc \au{{Bell}, A.~R.}} \yr{1978}  \at{{The acceleration of cosmic rays in
  shock fronts - I.}}  \jt{Mon. Not. R. Astron. Soc.}  \bvol{182},  \pg{147}.

\bibitem[Bender \& Orszag(1978)]{bender_book}
{\sc \au{Bender, C.~M.} \& \au{Orszag, S.~A.}} \yr{1978} {\em {Advanced
  Mathematical Methods for Scientists and Engineers}\/}.  \publ{McGraw-Hill}.

\bibitem[{Beraldo e Silva} {\em et~al.\/}(2017){Beraldo e Silva},
  de~Siqueira~Pedra, Sodr{\'{e}}, Perico \& Lima]{Beraldo_e_Silva_2017}
{\sc \au{{Beraldo e Silva}, L.}, \au{de~Siqueira~Pedra, W.}, \au{Sodr{\'{e}},
  L.}, \au{Perico, E. L.~D.} \& \au{Lima, M.}} \yr{2017}  \at{{The arrow of
  time in the collapse of collisionless self-gravitating systems: non-validity
  of the Vlasov{\textendash}Poisson equation during violent relaxation}}.
  \jt{Astrophys. J.}  \bvol{846},  \pg{125}.

\bibitem[Birdsall \& Langdon(1985)]{birdsall1985plasma}
{\sc \au{Birdsall, C.} \& \au{Langdon, A.}} \yr{1985} {\em Plasma Physics via
  Computer Simulation\/}.  \publ{CRC Press, United Kingdom}.

\bibitem[{Birn} {\em et~al.\/}(2012){Birn}, {Artemyev}, {Baker}, {Echim},
  {Hoshino} \& {Zelenyi}]{Birn_2012}
{\sc \au{{Birn}, J.}, \au{{Artemyev}, A.~V.}, \au{{Baker}, D.~N.}, \au{{Echim},
  M.}, \au{{Hoshino}, M.} \& \au{{Zelenyi}, L.~M.}} \yr{2012}  \at{{Particle
  acceleration in the magnetotail and aurora}}.  \jt{Space Sci. Rev.}
  \bvol{173},  \pg{49}.

\bibitem[Boltzmann(1896)]{Boltzmann}
{\sc \au{Boltzmann, L.}} \yr{1896} {\em Vorlesugnen \"uber Gastheorie\/}.
  \publ{Leipzig: J. A. Barth}.

\bibitem[Boris \& Shanny(1972)]{Boris_conferenceproc}
{\sc \au{Boris, J.} \& \au{Shanny, R.}} \yr{1972} {\em {Proceedings of the 4th
  Conference on Numerical Simulation of Plasmas}\/}.  \publ{Naval Research
  Laboratory, Washington}.

\bibitem[Caprioli \& Spitkovsky(2014)]{Caprioli_2014}
{\sc \au{Caprioli, D.} \& \au{Spitkovsky, A.}} \yr{2014}  \at{{Simulations of
  ion acceleration at non-relativistic shocks. i. acceleration efficiency}}.
  \jt{Astrophys. J.}  \bvol{783},  \pg{91}.

\bibitem[Chandran(2000)]{Chandran_2000}
{\sc \au{Chandran, B. D.~G.}} \yr{2000}  \at{{Scattering of energetic particles
  by anisotropic magnetohydrodynamic turbulence with a Goldreich-Sridhar power
  spectrum}}.  \jt{Phys. Rev. Lett.}  \bvol{85},  \pg{4656}.

\bibitem[Chavanis(2004)]{Chavanis2004}
{\sc \au{Chavanis, P.-H.}} \yr{2004}  \at{{Generalized thermodynamics and
  kinetic equations: Boltzmann, Landau, Kramers and Smoluchowski}}.
  \jt{Physica A}  \bvol{332},  \pg{89}.

\bibitem[Chavanis(2006{\natexlab{{\em a\/}}})]{Chavanis_2006b}
{\sc \au{Chavanis, P.-H.}} \yr{2006{\natexlab{{\em a\/}}}}  \at{Coarse-grained
  distributions and superstatistics}.  \jt{Physica A}  \bvol{359},  \pg{177}.

\bibitem[Chavanis(2006{\natexlab{{\em b\/}}})]{Chavanis_2006a}
{\sc \au{Chavanis, P.-H.}} \yr{2006{\natexlab{{\em b\/}}}}
  \at{Quasi-stationary states and incomplete violent relaxation in systems with
  long-range interactions}.  \jt{Physica A}  \bvol{365},  \pg{102}.

\bibitem[Chavanis(2022)]{Chavanis_2021}
{\sc \au{Chavanis, P.-H.}} \yr{2022}  \at{Kinetic theory of collisionless
  relaxation for systems with long-range interactions}.  \jt{Physica A}
  \bvol{606},  \pg{128089}.

\bibitem[Chavanis {\em et~al.\/}(1996)Chavanis, Sommeria \&
  Robert]{Chavanis_1996}
{\sc \au{Chavanis, P.~H.}, \au{Sommeria, J.} \& \au{Robert, R.}} \yr{1996}
  \at{Statistical mechanics of two-dimensional vortices and collisionless
  stellar systems}.  \jt{Astrophys. J.}  \bvol{471},  \pg{385}.

\bibitem[Chen {\em et~al.\/}(2020)Chen, Bale, Bonnell, Borovikov, Bowen,
  Burgess, Case, Chandran, de~Wit, Goetz {\em et~al.\/}]{chen2020evolution}
{\sc \au{Chen, C.}, \au{Bale, S.}, \au{Bonnell, J.}, \au{Borovikov, D.},
  \au{Bowen, T.}, \au{Burgess, D.}, \au{Case, A.}, \au{Chandran, B.},
  \au{de~Wit, T.~D.}, \au{Goetz, K.} \& \au{others}} \yr{2020}  \at{The
  evolution and role of solar wind turbulence in the inner heliosphere}.
  \jt{Astrophys. J. Suppl. Ser.}  \bvol{246},  \pg{53}.

\bibitem[Comisso \& Sironi(2018)]{Comisso_2018}
{\sc \au{Comisso, L.} \& \au{Sironi, L.}} \yr{2018}  \at{Particle acceleration
  in relativistic plasma turbulence}.  \jt{Phys. Rev. Lett.}  \bvol{121},
  \pg{255101}.

\bibitem[Comisso \& Sironi(2022)]{Comisso_2022}
{\sc \au{Comisso, L.} \& \au{Sironi, L.}} \yr{2022}  \at{Ion and electron
  acceleration in fully kinetic plasma turbulence}.  \jt{Astrophys. J. Lett.}
  \bvol{936},  \pg{L27}.

\bibitem[Crumley {\em et~al.\/}(2019)Crumley, Caprioli, Markoff \&
  Spitkovsky]{Crumley_2019}
{\sc \au{Crumley, P.}, \au{Caprioli, D.}, \au{Markoff, S.} \& \au{Spitkovsky,
  A.}} \yr{2019}  \at{{Kinetic simulations of mildly relativistic shocks
  {\textendash} i. particle acceleration in high Mach number shocks}}.
  \jt{Mon. Not. R. Astron. Soc.}  \bvol{485},  \pg{5105}.

\bibitem[Cruz {\em et~al.\/}(2018)Cruz, Albertazzi, Bamford, Bell, Cross,
  Fraschetti, Graham, Hara, Kozlowski, Kuramitsu, Lamb, Lebedev, Marques,
  Miniati, Morita, Oliver, Reville, Sakawa, Sarkar, Spindloe, Trines,
  Tzeferacos, Silva, Bingham, Koenig \& Gregori]{Rigby_2018}
{\sc \au{Cruz, F.}, \au{Albertazzi, B.}, \au{Bamford, R.}, \au{Bell, A.~R.},
  \au{Cross, J.~E.}, \au{Fraschetti, F.}, \au{Graham, P.}, \au{Hara, Y.},
  \au{Kozlowski, P.~M.}, \au{Kuramitsu, Y.}, \au{Lamb, D.~Q.}, \au{Lebedev,
  S.}, \au{Marques, J.~R.}, \au{Miniati, F.}, \au{Morita, T.}, \au{Oliver, M.},
  \au{Reville, B.}, \au{Sakawa, Y.}, \au{Sarkar, S.}, \au{Spindloe, C.},
  \au{Trines, R.}, \au{Tzeferacos, P.}, \au{Silva, L.~O.}, \au{Bingham, R.},
  \au{Koenig, M.} \& \au{Gregori, G.}} \yr{2018}  \at{{Electron acceleration by
  wave turbulence in a magnetized plasma}}.  \jt{Nat. Phys.}  \bvol{14},
  \pg{475}.

\bibitem[Davis {\em et~al.\/}(2023)Davis, Avaria, Bora, Jain, Moreno, Pavez \&
  Soto]{Davis_2023}
{\sc \au{Davis, S.}, \au{Avaria, G.}, \au{Bora, B.}, \au{Jain, J.}, \au{Moreno,
  J.}, \au{Pavez, C.} \& \au{Soto, L.}} \yr{2023} A derivation of the kappa
  distribution in non-equilibrium, steady-state plasmas,  \arxiv{arXiv:
  2304.13792}.

\bibitem[Dodin \& Fisch(2005)]{Dodin_2005}
{\sc \au{Dodin, I.} \& \au{Fisch, N.}} \yr{2005}  \at{{Variational formulation
  of the Gardner's restacking algorithm}}.  \jt{Phys. Lett. A}  \bvol{341},
  \pg{187}.

\bibitem[{Dud{\'\i}k} {\em et~al.\/}(2017){Dud{\'\i}k}, {Dzif{\caran
  c}{\'a}kov{\'a}}, {Meyer-Vernet}, {Del Zanna}, {Young}, {Giunta},
  {Sylwester}, {Sylwester}, {Oka}, {Mason}, {Vocks}, {Matteini}, {Krucker},
  {Williams} \& {Mackovjak}]{Dudik_2017}
{\sc \au{{Dud{\'\i}k}, J.}, \au{{Dzif{\caran c}{\'a}kov{\'a}}, E.},
  \au{{Meyer-Vernet}, N.}, \au{{Del Zanna}, G.}, \au{{Young}, P.~R.},
  \au{{Giunta}, A.}, \au{{Sylwester}, B.}, \au{{Sylwester}, J.}, \au{{Oka},
  M.}, \au{{Mason}, H.~E.}, \au{{Vocks}, C.}, \au{{Matteini}, L.},
  \au{{Krucker}, S.}, \au{{Williams}, D.~R.} \& \au{{Mackovjak}, {\caran S}.}}
  \yr{2017}  \at{{Nonequilibrium processes in the solar corona, transition
  region, flares, and solar wind}}.  \jt{Solar Phys.}  \bvol{292},  \pg{100}.

\bibitem[{Ergun} {\em et~al.\/}(2020){Ergun}, {Ahmadi}, {Kromyda}, {Schwartz},
  {Chasapis}, {Hoilijoki}, {Wilder}, {Cassak}, {Stawarz}, {Goodrich}, {Turner},
  {Pucci}, {Pouquet}, {Matthaeus}, {Drake}, {Hesse}, {Shay}, {Torbert} \&
  {Burch}]{Ergun_2020}
{\sc \au{{Ergun}, R.~E.}, \au{{Ahmadi}, N.}, \au{{Kromyda}, L.},
  \au{{Schwartz}, S.~J.}, \au{{Chasapis}, A.}, \au{{Hoilijoki}, S.},
  \au{{Wilder}, F.~D.}, \au{{Cassak}, P.~A.}, \au{{Stawarz}, J.~E.},
  \au{{Goodrich}, K.~A.}, \au{{Turner}, D.~L.}, \au{{Pucci}, F.},
  \au{{Pouquet}, A.}, \au{{Matthaeus}, W.~H.}, \au{{Drake}, J.~F.},
  \au{{Hesse}, M.}, \au{{Shay}, M.~A.}, \au{{Torbert}, R.~B.} \& \au{{Burch},
  J.~L.}} \yr{2020}  \at{{Particle acceleration in strong turbulence in the
  Earth's magnetotail}}.  \jt{Astrophys. J.}  \bvol{898},  \pg{153}.

\bibitem[Ewart {\em et~al.\/}(2022)Ewart, Brown, Adkins \&
  Schekochihin]{Ewart_2022}
{\sc \au{Ewart, R.}, \au{Brown, A.}, \au{Adkins, T.} \& \au{Schekochihin, A.}}
  \yr{2022}  \at{{Collisionless relaxation of a Lynden-Bell plasma}}.  \jt{J.
  Plasma Phys.}  \bvol{88},  \pg{925880501}.

\bibitem[Ferri{\`{e}}re(2019)]{Ferriere_2019}
{\sc \au{Ferri{\`{e}}re, K.}} \yr{2019}  \at{Plasma turbulence in the
  interstellar medium}.  \jt{Plasma Phys. Control. Fusion}  \bvol{62},
  \pg{014014}.

\bibitem[Fisk \& Gloeckler(2014)]{Fisk_2014}
{\sc \au{Fisk, L.~A.} \& \au{Gloeckler, G.}} \yr{2014}  \at{The case for a
  common spectrum of particles accelerated in the heliosphere: Observations and
  theory}.  \jt{J. Geophys. Res. Space Phys.}  \bvol{119},  \pg{8733}.

\bibitem[Gardner(1963)]{Gardner63}
{\sc \au{Gardner, C.~S.}} \yr{1963}  \at{Bound on the energy available from a
  plasma}.  \jt{Phys. Fluids}  \bvol{6},  \pg{839}.

\bibitem[Gloeckler {\em et~al.\/}(2008)Gloeckler, Fisk, Mason \&
  Hill]{Gloeckler2008}
{\sc \au{Gloeckler, G.}, \au{Fisk, L.~A.}, \au{Mason, G.~M.} \& \au{Hill,
  M.~E.}} \yr{2008}  \at{Formation of power law tail with spectral index‐5
  inside and beyond the heliosphere}.  \jt{AIP Conf. Proc.}  \bvol{1039},
  \pg{367}.

\bibitem[Hartouni {\em et~al.\/}(2022)Hartouni, Moore, Crilly, Appelbe, Amendt,
  Baker, Casey, Clark, Döppner, Eckart, Field, Gatu-Johnson, Grim, Hatarik,
  Jeet, Kerr, Kilkenny, Kritcher, Meaney \& Zylstra]{Hartouni_2022}
{\sc \au{Hartouni, E.}, \au{Moore, A.}, \au{Crilly, A.}, \au{Appelbe, B.},
  \au{Amendt, P.}, \au{Baker, K.}, \au{Casey, D.}, \au{Clark, D.},
  \au{Döppner, T.}, \au{Eckart, M.}, \au{Field, J.}, \au{Gatu-Johnson, M.},
  \au{Grim, G.}, \au{Hatarik, R.}, \au{Jeet, J.}, \au{Kerr, S.}, \au{Kilkenny,
  J.}, \au{Kritcher, A.}, \au{Meaney, K.} \& \au{Zylstra, A.}} \yr{2022}
  \at{Evidence for suprathermal ion distribution in burning plasmas}.  \jt{Nat.
  Phys.}  \bvol{19},  \pg{1}.

\bibitem[Havrda \& Charv{\'a}t(1967)]{Havrda1967}
{\sc \au{Havrda, J.} \& \au{Charv{\'a}t, F.}} \yr{1967}  \at{{Quantification
  method of classification processes. Concept of structural
  {$\alpha$}-entropy}}.  \jt{Kybernetika}  \bvol{3},  \pg{30}.

\bibitem[Helander(2017)]{helander_2017}
{\sc \au{Helander, P.}} \yr{2017}  \at{Available energy and ground states of
  collisionless plasmas}.  \jt{J. Plasma Phys.}  \bvol{83},  \pg{715830401}.

\bibitem[Helander(2020)]{helander_2020}
{\sc \au{Helander, P.}} \yr{2020}  \at{Available energy of magnetically
  confined plasmas}.  \jt{J. Plasma Phys.}  \bvol{86},  \pg{905860201}.

\bibitem[Helander \& Sigmar(2005)]{Helander_book}
{\sc \au{Helander, P.} \& \au{Sigmar, D.~J.}} \yr{2005} {\em Collisional
  Transport in Magnetised Plasmas\/}.  \publ{Cambridge: Cambridge University
  Press}.

\bibitem[Hopkins {\em et~al.\/}(2022)Hopkins, Squire, Butsky \&
  Ji]{Hopkins_2022}
{\sc \au{Hopkins, P.~F.}, \au{Squire, J.}, \au{Butsky, I.~S.} \& \au{Ji, S.}}
  \yr{2022}  \at{Standard self-confinement and extrinsic turbulence models for
  cosmic ray transport are fundamentally incompatible with observations}.
  \jt{Mon. Not. R. Astron. Soc.}  \bvol{517},  \pg{5413}.

\bibitem[Kadomtsev \& Pogutse(1970)]{Kadomtsev_Pogutse70}
{\sc \au{Kadomtsev, B.~B.} \& \au{Pogutse, O.~P.}} \yr{1970}
  \at{{Collisionless relaxation in systems with Coulomb interactions}}.
  \jt{Phys. Rev. Lett.}  \bvol{25},  \pg{1155}.

\bibitem[Kawazura {\em et~al.\/}(2022)Kawazura, Schekochihin, Barnes, Dorland
  \& Balbus]{kawazura_schekochihin_barnes_dorland_balbus_2022}
{\sc \au{Kawazura, Y.}, \au{Schekochihin, A.}, \au{Barnes, M.}, \au{Dorland,
  W.} \& \au{Balbus, S.}} \yr{2022}  \at{{Energy partition between Alfvénic
  and compressive fluctuations in magnetorotational turbulence with
  near-azimuthal mean magnetic field}}.  \jt{J. Plasma Phys.}  \bvol{88},
  \pg{905880311}.

\bibitem[Kempski \& Quataert(2022)]{Kempski_2022}
{\sc \au{Kempski, P.} \& \au{Quataert, E.}} \yr{2022}  \at{{Reconciling cosmic
  ray transport theory with phenomenological models motivated by Milky-Way
  data}}.  \jt{Mon. Not. R. Astron. Soc.}  \bvol{514},  \pg{657}.

\bibitem[Kolmes \& Fisch(2020)]{Kolmes_2020b}
{\sc \au{Kolmes, E.~J.} \& \au{Fisch, N.~J.}} \yr{2020}  \at{{Recovering
  Gardner restacking with purely diffusive operations}}.  \jt{Phys. Rev. E}
  \bvol{102},  \pg{063209}.

\bibitem[Kolmes {\em et~al.\/}(2020)Kolmes, Helander \& Fisch]{Kolmes_2020a}
{\sc \au{Kolmes, E.~J.}, \au{Helander, P.} \& \au{Fisch, N.~J.}} \yr{2020}
  \at{Available energy from diffusive and reversible phase space
  rearrangements}.  \jt{Phys. Plasmas}  \bvol{27},  \pg{062110}.

\bibitem[Kunz {\em et~al.\/}(2016)Kunz, Stone \& Quataert]{Kunz_2016}
{\sc \au{Kunz, M.~W.}, \au{Stone, J.~M.} \& \au{Quataert, E.}} \yr{2016}
  \at{{Magnetorotational turbulence and dynamo in a collisionless plasma}}.
  \jt{Phys. Rev. Lett.}  \bvol{117},  \pg{235101}.

\bibitem[Landau(1936)]{Landau1936}
{\sc \au{Landau, L.~D.}} \yr{1936}  \at{{Transport equation in the case of
  Coulomb interaction}}.  \jt{Zh. Eksp. Teor. Fiz.}  \bvol{7},  \pg{203}.

\bibitem[Lenard(1960)]{Lenard60}
{\sc \au{Lenard, A.}} \yr{1960}  \at{{On Bogoliubov's kinetic equation for a
  spatially homogeneous plasma}}.  \jt{Ann. Phys.}  \bvol{10},  \pg{390}.

\bibitem[Lesur(2021)]{lesur_2021}
{\sc \au{Lesur, G. R.~J.}} \yr{2021}  \at{Magnetohydrodynamics of
  protoplanetary discs}.  \jt{J. Plasma Phys.}  \bvol{87},  \pg{205870101}.

\bibitem[Levin {\em et~al.\/}(2014)Levin, Pakter, Rizzato, Teles \&
  Benetti]{Levin_2014}
{\sc \au{Levin, Y.}, \au{Pakter, R.}, \au{Rizzato, F.~B.}, \au{Teles, T.~N.} \&
  \au{Benetti, F.~P.}} \yr{2014}  \at{Nonequilibrium statistical mechanics of
  systems with long-range interactions}.  \jt{Phys. Rep.}  \bvol{535},  \pg{1}.

\bibitem[Levin {\em et~al.\/}(2008)Levin, Pakter \& Teles]{Levin_2008}
{\sc \au{Levin, Y.}, \au{Pakter, R.} \& \au{Teles, T.~N.}} \yr{2008}
  \at{Collisionless relaxation in non-neutral plasmas}.  \jt{Phys. Rev. Lett.}
  \bvol{100},  \pg{040604}.

\bibitem[{Livadiotis} {\em et~al.\/}(2018){Livadiotis}, {Desai} \& {Wilson
  III}]{Livadiotis_2018}
{\sc \au{{Livadiotis}, G.}, \au{{Desai}, M.~I.} \& \au{{Wilson III}, L.~B.}}
  \yr{2018}  \at{{Generation of kappa distributions in solar wind at 1 au}}.
  \jt{Astrophys. J.}  \bvol{853},  \pg{142}.

\bibitem[Livadiotis \& McComas(2009)]{Livadiotis_2009}
{\sc \au{Livadiotis, G.} \& \au{McComas, D.}} \yr{2009}  \at{{Beyond kappa
  distributions: exploiting Tsallis statistical mechanics in space plasmas}}.
  \jt{J. Geophys. Res.}  \bvol{114},  \pg{A11105}.

\bibitem[Lucek {\em et~al.\/}(2005)Lucek, Constantinescu, Goldstein, Pickett,
  Pincon, Sahraoui, Treumann \& Walker]{lucek2005magnetosheath}
{\sc \au{Lucek, E.}, \au{Constantinescu, D.}, \au{Goldstein, M.}, \au{Pickett,
  J.}, \au{Pincon, J.-L.}, \au{Sahraoui, F.}, \au{Treumann, R.} \& \au{Walker,
  S.}} \yr{2005}  \at{The magnetosheath}.  \jt{Space Sci. Rev.}  \bvol{118},
  \pg{95}.

\bibitem[Lynden-Bell(1967)]{LyndenBell67}
{\sc \au{Lynden-Bell, D.}} \yr{1967}  \at{Statistical mechanics of violent
  relaxation in stellar systems}.  \jt{Mon. Not. R. Astron. Soc.}  \bvol{136},
  \pg{101}.

\bibitem[Mackenbach {\em et~al.\/}(2022)Mackenbach, Proll \&
  Helander]{MackenBach_2022}
{\sc \au{Mackenbach, R. J.~J.}, \au{Proll, J. H.~E.} \& \au{Helander, P.}}
  \yr{2022}  \at{Available energy of trapped electrons and its relation to
  turbulent transport}.  \jt{Phys. Rev. Lett.}  \bvol{128},  \pg{{175001}}.

\bibitem[Mailloux {\em et~al.\/}(2022)]{Mailloux_2022}
{\sc \au{Mailloux, J.} \& \au{others}} \yr{2022}  \at{{Overview of JET results
  for optimising ITER operation}}.  \jt{Nucl. Fusion}  \bvol{62},  \pg{042026}.

\bibitem[Maxwell(1860)]{Maxwell1860}
{\sc \au{Maxwell, J.~C.}} \yr{1860}  \at{{V. Illustrations of the dynamical
  theory of gases.—Part I. On the motions and collisions of perfectly elastic
  spheres}}.  \jt{The London, Edinburgh, and Dublin Philosophical Magazine and
  Journal of Science}  \bvol{19},  \pg{19}.

\bibitem[May {\em et~al.\/}(2014)May, Tonge, Ellis, Mori, Fiuza, Fonseca, Silva
  \& Ren]{May_2014}
{\sc \au{May, J.}, \au{Tonge, J.}, \au{Ellis, I.}, \au{Mori, W.~B.}, \au{Fiuza,
  F.}, \au{Fonseca, R.~A.}, \au{Silva, L.~O.} \& \au{Ren, C.}} \yr{2014}
  \at{{Enhanced stopping of macro-particles in particle-in-cell simulations}}.
  \jt{Phys. Plasmas}  \bvol{21},  \pg{052703}.

\bibitem[Moncuquet {\em et~al.\/}(2020)Moncuquet, Meyer-Vernet, Issautier,
  Pulupa, Bonnell, Bale, de~Wit, Goetz, Griton, Harvey, MacDowall, Maksimovic
  \& Malaspina]{Moncuquet_2020}
{\sc \au{Moncuquet, M.}, \au{Meyer-Vernet, N.}, \au{Issautier, K.}, \au{Pulupa,
  M.}, \au{Bonnell, J.~W.}, \au{Bale, S.~D.}, \au{de~Wit, T.~D.}, \au{Goetz,
  K.}, \au{Griton, L.}, \au{Harvey, P.~R.}, \au{MacDowall, R.~J.},
  \au{Maksimovic, M.} \& \au{Malaspina, D.~M.}} \yr{2020}  \at{{First in situ
  measurements of electron density and temperature from quasi-thermal noise
  spectroscopy with Parker Solar Probe/{FIELDS}}}.  \jt{Astrophys. J. Suppl.
  Ser.}  \bvol{246},  \pg{44}.

\bibitem[Mora \& Bialek(2011)]{Mora_2011}
{\sc \au{Mora, T.} \& \au{Bialek, W.}} \yr{2011}  \at{Are biological systems
  poised at criticality?}  \jt{J. Stat. Phys.}  \bvol{144}~(2),  \pg{268}.

\bibitem[Nastac {\em et~al.\/}(2023)Nastac, Ewart, Sengupta, Schekochihin,
  Barnes \& Dorland]{Nastac_2022}
{\sc \au{Nastac, M.~L.}, \au{Ewart, R.~J.}, \au{Sengupta, W.},
  \au{Schekochihin, A.~A.}, \au{Barnes, M.} \& \au{Dorland, W.}} \yr{2023}
  {Phase-space entropy cascade and irreversibility of stochastic heating in
  nearly collisionless plasma turbulence. In preparation}.

\bibitem[Oka {\em et~al.\/}(2018)Oka, Birn, Battaglia, Chaston, Hatch,
  Livadiotis, Imada, Miyoshi, Kuhar, Effenberger, Eriksson, Khotyaintsev \&
  Retin{\`{o}}]{Oka_2018}
{\sc \au{Oka, M.}, \au{Birn, J.}, \au{Battaglia, M.}, \au{Chaston, C.~C.},
  \au{Hatch, S.~M.}, \au{Livadiotis, G.}, \au{Imada, S.}, \au{Miyoshi, Y.},
  \au{Kuhar, M.}, \au{Effenberger, F.}, \au{Eriksson, E.}, \au{Khotyaintsev,
  Y.~V.} \& \au{Retin{\`{o}}, A.}} \yr{2018}  \at{{Electron power-law spectra
  in solar and space plasmas}}.  \jt{Space Sci. Rev.}  \bvol{214}.

\bibitem[{Oka} {\em et~al.\/}(2015){Oka}, {Krucker}, {Hudson} \&
  {Saint-Hilaire}]{Oka_2015}
{\sc \au{{Oka}, M.}, \au{{Krucker}, S.}, \au{{Hudson}, H.~S.} \&
  \au{{Saint-Hilaire}, P.}} \yr{2015}  \at{{Electron energy partition in the
  above-the-looptop solar hard X-ray sources}}.  \jt{Astrophys. J.}
  \bvol{799},  \pg{129}.

\bibitem[{Ormes} \& {Freier}(1978)]{Ormes_1978}
{\sc \au{{Ormes}, J.} \& \au{{Freier}, P.}} \yr{1978}  \at{{On the propagation
  of cosmic rays in the galaxy.}}  \jt{Astrophys. J.}  \bvol{222},  \pg{471}.

\bibitem[Pierrard \& Lazar(2010)]{Pierrard_2010}
{\sc \au{Pierrard, V.} \& \au{Lazar, M.}} \yr{2010}  \at{{Kappa distributions:
  theory and applications in space plasmas}}.  \jt{Sol. Phys.}  \bvol{267},
  \pg{153}.

\bibitem[Reichherzer {\em et~al.\/}(2021)Reichherzer, Merten, Dörner, Tjus,
  Pueschel \& Zweibel]{Reichherzer_2021}
{\sc \au{Reichherzer, P.}, \au{Merten, L.}, \au{Dörner, J.}, \au{Tjus, J.~B.},
  \au{Pueschel, M.~J.} \& \au{Zweibel, E.~G.}} \yr{2021}  \at{Regimes of
  cosmic-ray diffusion in galactic turbulence}.  \jt{{SN} Appl. Sci.}
  \bvol{4}.

\bibitem[Robert \& Sommeria(1991)]{Robert_Sommeria_1991}
{\sc \au{Robert, R.} \& \au{Sommeria, J.}} \yr{1991}  \at{Statistical
  equilibrium states for two-dimensional flows}.  \jt{J. Fluid Mech.}
  \bvol{229},  \pg{291}.

\bibitem[Ruszkowski \& Pfrommer(2023)]{ruszkowski2023cosmic}
{\sc \au{Ruszkowski, M.} \& \au{Pfrommer, C.}} \yr{2023} Cosmic ray feedback in
  galaxies and galaxy clusters -- a pedagogical introduction and a topical
  review of the acceleration, transport, observables, and dynamical impact of
  cosmic rays,  \arxiv{arXiv: 2306.03141}.

\bibitem[Schekochihin {\em et~al.\/}(2009)Schekochihin, Cowley, Dorland,
  Hammett, Howes, Quataert \& Tatsuno]{Schekochihin_2009}
{\sc \au{Schekochihin, A.~A.}, \au{Cowley, S.~C.}, \au{Dorland, W.},
  \au{Hammett, G.~W.}, \au{Howes, G.~G.}, \au{Quataert, E.} \& \au{Tatsuno,
  T.}} \yr{2009}  \at{Astrophysical gyrokinetics: kinetic and fluid turbulent
  cascades in magnetized weakly collisional plasmas}.  \jt{Astrophys. J. Suppl.
  Ser.}  \bvol{182},  \pg{310}.

\bibitem[{Schlickeiser}(1989)]{Schlickeiser_1989}
{\sc \au{{Schlickeiser}, R.}} \yr{1989}  \at{{Cosmic-ray transport and
  acceleration. i. derivation of the kinetic equation and application to cosmic
  rays in static cold media}}.  \jt{Astrophys. J.}  \bvol{336},  \pg{243}.

\bibitem[Schwab {\em et~al.\/}(2014)Schwab, Nemenman \& Mehta]{Schwab_2014}
{\sc \au{Schwab, D.~J.}, \au{Nemenman, I.} \& \au{Mehta, P.}} \yr{2014}
  \at{Zipf's law and criticality in multivariate data without fine-tuning}.
  \jt{Phys. Rev. Lett.}  \bvol{113},  \pg{068102}.

\bibitem[Severne \& Luwel(1980)]{Severne_1980}
{\sc \au{Severne, G.} \& \au{Luwel, M.}} \yr{1980}  \at{{Dynamical theory of
  collisionless relaxation}}.  \jt{Astrophys. Space Sci.}  \bvol{72},
  \pg{293}.

\bibitem[Sironi \& Spitkovsky(2010)]{Sironi_2010}
{\sc \au{Sironi, L.} \& \au{Spitkovsky, A.}} \yr{2010}  \at{Particle
  acceleration in relativistic magnetized collisionless electron-ion shocks}.
  \jt{Astrophys. J.}  \bvol{726},  \pg{75}.

\bibitem[Sironi \& Spitkovsky(2014)]{Sironi_2014}
{\sc \au{Sironi, L.} \& \au{Spitkovsky, A.}} \yr{2014}  \at{{Relativistic
  reconnection: an efficient source of non-thermal particles}}.  \jt{Astrophys.
  J.}  \bvol{783},  \pg{L21}.

\bibitem[Su \& Oberman(1968)]{SuOberman1968}
{\sc \au{Su, C.~H.} \& \au{Oberman, C.}} \yr{1968}  \at{Collisional damping of
  a plasma echo}.  \jt{Phys. Rev. Lett.}  \bvol{20},  \pg{427}.

\bibitem[Touati {\em et~al.\/}(2022)Touati, Codur, Tsung, Decyk, Mori \&
  Silva]{Touati_2022}
{\sc \au{Touati, M.}, \au{Codur, R.}, \au{Tsung, F.}, \au{Decyk, V.~K.},
  \au{Mori, W.~B.} \& \au{Silva, L.~O.}} \yr{2022}  \at{{Kinetic theory of
  particle-in-cell simulation plasma and the ensemble averaging technique}}.
  \jt{Plasma Phys. Control. Fusion}  \bvol{64},  \pg{115014}.

\bibitem[{Tsallis}(1988)]{Tsallis1988}
{\sc \au{{Tsallis}, C.}} \yr{1988}  \at{{Possible generalization of
  Boltzmann-Gibbs statistics}}.  \jt{J. Stat. Phys.}  \bvol{52},  \pg{479}.

\bibitem[Uzdensky(2022)]{uzdensky_2022}
{\sc \au{Uzdensky, D.~A.}} \yr{2022}  \at{{Relativistic non-thermal particle
  acceleration in two-dimensional collisionless magnetic reconnection}}.
  \jt{J. Plasma Phys.}  \bvol{88},  \pg{905880114}.

\bibitem[Verscharen {\em et~al.\/}(2019)Verscharen, Klein \&
  Maruca]{Verscharen_2019}
{\sc \au{Verscharen, D.}, \au{Klein, K.~G.} \& \au{Maruca, B.~A.}} \yr{2019}
  \at{The multi-scale nature of the solar wind}.  \jt{Living Rev. Sol. Phys.}
  \bvol{16},  \pg{5}.

\bibitem[Werner \& Uzdensky(2017)]{Werner_2017}
{\sc \au{Werner, G.~R.} \& \au{Uzdensky, D.~A.}} \yr{2017}  \at{{Nonthermal
  particle acceleration in 3D relativistic magnetic reconnection in pair
  plasma}}.  \jt{Astrophys. J. Lett.}  \bvol{843},  \pg{L27}.

\bibitem[Werner \& Uzdensky(2021)]{werner_uzdensky_2021}
{\sc \au{Werner, G.~R.} \& \au{Uzdensky, D.~A.}} \yr{2021}  \at{{Reconnection
  and particle acceleration in three-dimensional current sheet evolution in
  moderately magnetized astrophysical pair plasma}}.  \jt{J. Plasma Phys.}
  \bvol{87},  \pg{905870613}.

\bibitem[Yang {\em et~al.\/}(2020)Yang, Wang, Zhao, Tao, Li,
  Wimmer-Schweingruber, He, Tian \& Bale]{Yang_2020}
{\sc \au{Yang, L.}, \au{Wang, L.}, \au{Zhao, L.}, \au{Tao, J.}, \au{Li, G.},
  \au{Wimmer-Schweingruber, R.~F.}, \au{He, J.}, \au{Tian, H.} \& \au{Bale,
  S.~D.}} \yr{2020}  \at{{Quiet-time solar wind suprathermal electrons of
  different solar origins}}.  \jt{Astrophys. J. Lett.}  \bvol{896},  \pg{L5}.

\bibitem[Zhdankin(2021)]{Zhdankin_2021b}
{\sc \au{Zhdankin, V.}} \yr{2021}  \at{{Particle energization in relativistic
  plasma turbulence: solenoidal versus compressive driving}}.  \jt{Astrophys.
  J.}  \bvol{922},  \pg{172}.

\bibitem[Zhdankin(2022{\natexlab{{\em a\/}}})]{Zhdankin_2021a}
{\sc \au{Zhdankin, V.}} \yr{2022{\natexlab{{\em a\/}}}}  \at{Generalized
  entropy production in collisionless plasma flows and turbulence}.  \jt{Phys.
  Rev. X}  \bvol{12},  \pg{031011}.

\bibitem[Zhdankin(2022{\natexlab{{\em b\/}}})]{Zhdankin_2022}
{\sc \au{Zhdankin, V.}} \yr{2022{\natexlab{{\em b\/}}}}  \at{Non-thermal
  particle acceleration from maximum entropy in collisionless plasmas}.  \jt{J.
  Plasma Phys.}  \bvol{88},  \pg{175880303}.

\bibitem[Zhdankin {\em et~al.\/}(2019)Zhdankin, Uzdensky, Werner \&
  Begelman]{Zhdankin_2019}
{\sc \au{Zhdankin, V.}, \au{Uzdensky, D.~A.}, \au{Werner, G.~R.} \&
  \au{Begelman, M.~C.}} \yr{2019}  \at{{Electron and ion energization in
  relativistic plasma turbulence}}.  \jt{Phys. Rev. Lett.}  \bvol{122},
  \pg{055101}.

\bibitem[Zhdankin {\em et~al.\/}(2017)Zhdankin, Werner, Uzdensky \&
  Begelman]{Zhdankin_2017}
{\sc \au{Zhdankin, V.}, \au{Werner, G.~R.}, \au{Uzdensky, D.~A.} \&
  \au{Begelman, M.~C.}} \yr{2017}  \at{Kinetic turbulence in relativistic
  plasma: from thermal bath to nonthermal continuum}.  \jt{Phys. Rev. Lett.}
  \bvol{118},  \pg{055103}.

\bibitem[Zhuravleva {\em et~al.\/}(2014)Zhuravleva, Churazov, Schekochihin,
  Allen, Ar{\'{e}}valo, Fabian, Forman, Sanders, Simionescu, Sunyaev, Vikhlinin
  \& Werner]{Zhuravleva_2014}
{\sc \au{Zhuravleva, I.}, \au{Churazov, E.}, \au{Schekochihin, A.~A.},
  \au{Allen, S.~W.}, \au{Ar{\'{e}}valo, P.}, \au{Fabian, A.~C.}, \au{Forman,
  W.~R.}, \au{Sanders, J.~S.}, \au{Simionescu, A.}, \au{Sunyaev, R.},
  \au{Vikhlinin, A.} \& \au{Werner, N.}} \yr{2014}  \at{{Turbulent heating in
  galaxy clusters brightest in X-rays}}.  \jt{Nature}  \bvol{515},  \pg{85}.

\end{thebibliography}

\end{document}